\documentclass[preprint,
superscriptaddress,
showpacs,
 amsmath,amssymb,
 aps,
pra,
latexsym,mathtools
]{revtex4-1}

\usepackage{graphicx}
\usepackage{dcolumn}
\usepackage{bm}
\usepackage{color}
\usepackage[bottom]{footmisc}
\usepackage{ulem}
\usepackage{setspace}
\usepackage[nottoc]{tocbibind}
\usepackage{blindtext}
\usepackage{titlesec}
\usepackage{cancel}

\usepackage[mathlines]{lineno}

\begin{document}


\title{Stabilization of Kerr-cat qubits with quantum circuit refrigerator}

\author{Shumpei Masuda}
\affiliation{Global Research and Development Center for Business by Quantum-AI technology (G-QuAT), National Institute of Advanced Industrial Science and Technology (AIST), 1-1-1, Umezono, Tsukuba, Ibaraki 305-8568, Japan}
\affiliation{NEC-AIST Quantum Technology Cooperative Research Laboratory, National Institute of Advanced Industrial Science and Technology (AIST), 1-1-1, Umezono, Tsukuba, Ibaraki 305-8568, Japan}
\email{shumpei.masuda@aist.go.jp}
\author{Shunsuke Kamimura}
\affiliation{Global Research and Development Center for Business by Quantum-AI technology (G-QuAT), National Institute of Advanced Industrial Science and Technology (AIST), 1-1-1, Umezono, Tsukuba, Ibaraki 305-8568, Japan}
\author{Tsuyoshi Yamamoto}
\affiliation{Faculty of Pure and Applied Sciences, University of Tsukuba, Tsukuba, Ibaraki 305-8571, Japan.}
\author{Takaaki Aoki}
\affiliation{Global Research and Development Center for Business by Quantum-AI technology (G-QuAT), National Institute of Advanced Industrial Science and Technology (AIST), 1-1-1, Umezono, Tsukuba, Ibaraki 305-8568, Japan}
\author{Akiyoshi Tomonaga}
\affiliation{Global Research and Development Center for Business by Quantum-AI technology (G-QuAT), National Institute of Advanced Industrial Science and Technology (AIST), 1-1-1, Umezono, Tsukuba, Ibaraki 305-8568, Japan}
\affiliation{NEC-AIST Quantum Technology Cooperative Research Laboratory, National Institute of Advanced Industrial Science and Technology (AIST), 1-1-1, Umezono, Tsukuba, Ibaraki 305-8568, Japan}

\date{\today}

\begin{abstract}
A periodically-driven superconducting nonlinear resonator can implement a Kerr-cat qubit, which provides a promising route to a quantum computer with a long lifetime. However, the system is vulnerable to pure dephasing, which causes unwanted excitations outside the qubit subspace. Therefore, we require a refrigeration technology which confines the system in the qubit subspace. We theoretically study on-chip refrigeration for Kerr-cat qubits based on photon-assisted electron tunneling at tunneling junctions, called quantum circuit refrigerator (QCR). Rates of QCR-induced deexcitations of the system can be changed by more than four orders of magnitude by tuning a bias voltage across the tunneling junctions. Unwanted QCR-induced bit flips are greatly suppressed due to quantum interference in the tunneling process, and thus the long lifetime is preserved. The QCR can serve as a tunable dissipation source which stabilizes Kerr-cat qubits.
\end{abstract}

\maketitle



\section*{INTRODUCTION}

In the quantum regime, where the nonlinearity is greater than the photon loss rate, a periodically-driven superconducting nonlinear resonator can operate as a Kerr-cat qubit, providing a promising route to a quantum computer~\cite{Milburn1991,Wielinga1993,Cochrane1999,Goto2016,Puri2017b}.
For example, a Kerr-cat qubit can be implemented by a parametrically-driven superconducting resonator with the Kerr nonlinearity, Kerr parametric oscillator (KPO)~\cite{Meaney2014,Wang2019,Goto2019}, and a superconducting nonlinear asymmetric inductive element (SNAIL)~\cite{Grimm2020,He2023}.
Two meta-stable states of the driven superconducting resonators are utilized as a Kerr-cat qubit.
Because of their long lifetime, the bit-flip error is much smaller than the phase-flip error. 
This biased nature of errors enables us to perform quantum error corrections with less overhead compared to other qubits with unbiased errors~\cite{Tuckett2019,Ataides2021}.
Qubit-gate operations have been intensely studied~\cite{Goto2016,Puri2017b,Puri2020,Kanao2022,Masuda2022,Chono2022,Aoki2023} and demonstrated~\cite{Grimm2020,Iyama2023}, and high error-correction performance by concatenating the XZZX surface code~\cite{Ataides2021} with Kerr-cat qubits~\cite{Darmawan2021} was examined.
Kerr-cat qubits also find applications to quantum annealing \cite{Goto2016,Nigg2017,Puri2017,Zhao2018,Onodera2020,Goto2020a,Kewming2020,Kanao2021,Yamaji2022}
and Boltzmann sampling~\cite{Goto2018}, and offer a platform to study quantum phase transitions~\cite{Dykman2018,Rota2019,Kewming2022} and quantum chaos~\cite{Milburn1991,Hovsepyan2016,Goto2021b}.

Kerr-cat qubits are vulnerable to pure dephasing of the resonator, which causes excitations outside the qubit subspace.
Such leakage errors into excited states are hard to correct by quantum error correction protocols, which only deal with errors in the qubit subspace. 
Although engineered two-photon loss realized with the help of a dissipative mode~\cite{Touzard2018,Puri2020} and frequency-selective dissipation~\cite{Putterman2022} can mitigate unwanted excitations, no experimental demonstration with a KPO or a SNAIL has been reported.
On the other hand, external single-photon loss in the resonator can reduce the excitations by transferring the population from excited states to the qubit subspace, but it generates phase-flips of the Kerr-cat qubit (errors in the qubit subspace)~\cite{Puri2017}.
Therefore, the photon loss rate needs to be small while still being large enough to mitigate the effects of the pure dephasing.
Since it is difficult to determine the amplitude of the pure dephasing before measurement, and the amplitude depends on the parameters used for an operation of the Kerr-cat qubit, making the photon loss tunable is a solution to meet the above requirement.

The electron tunneling through a microscopic junction can occur accompanied by energy exchange with the environment~\cite{Tien1963,Devoret1990,Girvin1990,Averin1990,Pekola2010}.
Thus, controlled electron tunneling offers a way to selectively cool or heat devices in a cryogenic refrigerator~\cite{Lowell2013}.
Remarkably, on-chip refrigeration of a superconducting resonator, which is based on the photon-assisted electron tunneling through normal-metal--insulator--superconductor (NIS) junctions, was demonstrated ~\cite{Tan2017,Silveri2019}.
This device, called quantum circuit refrigerator (QCR), can operate as a tunable source of dissipation for 
quantum-electric devices such as qubits~\cite{Silveri2017,Hsu2020,Hsu2021,Yoshioka2021}.
More recently, the qubit reset with QCR was demonstrated~\cite{Sevriuk2022,Yoshioka2023}. 
While several works on the effect of the QCR on linear resonators, two-level systems and transmons have been reported, it remains largely unexplored for periodically-driven systems such as KPOs.
Although it is intuitively expected that the QCR can cool KPOs more or less, it is nontrivial how the following characteristics of KPOs affect the effectiveness of the QCR:
(i) the system is periodically driven and its effective energy eigenstates exist in a rotating frame; 
(ii) the system typically has degenerate energy eigenstates with the biased errors.
Another important question is whether the biased nature of errors of KPOs will be preserved under the operation of the QCR.


We study the effect of electron tunneling through microscopic junctions on a periodically-driven superconducting resonator, particularly considering a \textcolor{black}{superconductor--insulator--normal-metal--insulator--superconductor (SINIS)} junction coupled to a KPO.
We develop the master equation which can describe the effect of the electron tunnelings on the coherence of the KPO as well as the inter-level population transfers.
Performance of the cooling based on the QCR is examined with the master equation. 
We also study drawbacks of the QCR such as the QCR-induced phase flip and bit flip.
We show that the QCR-induced bit flip is suppressed by more than six orders of magnitude when relevant energy levels of the KPO are degenerate due to quantum interference of the tunneling processes, and thus the biased nature of errors is preserved. 

\begin{figure}
\begin{center}
\includegraphics[width=15cm]{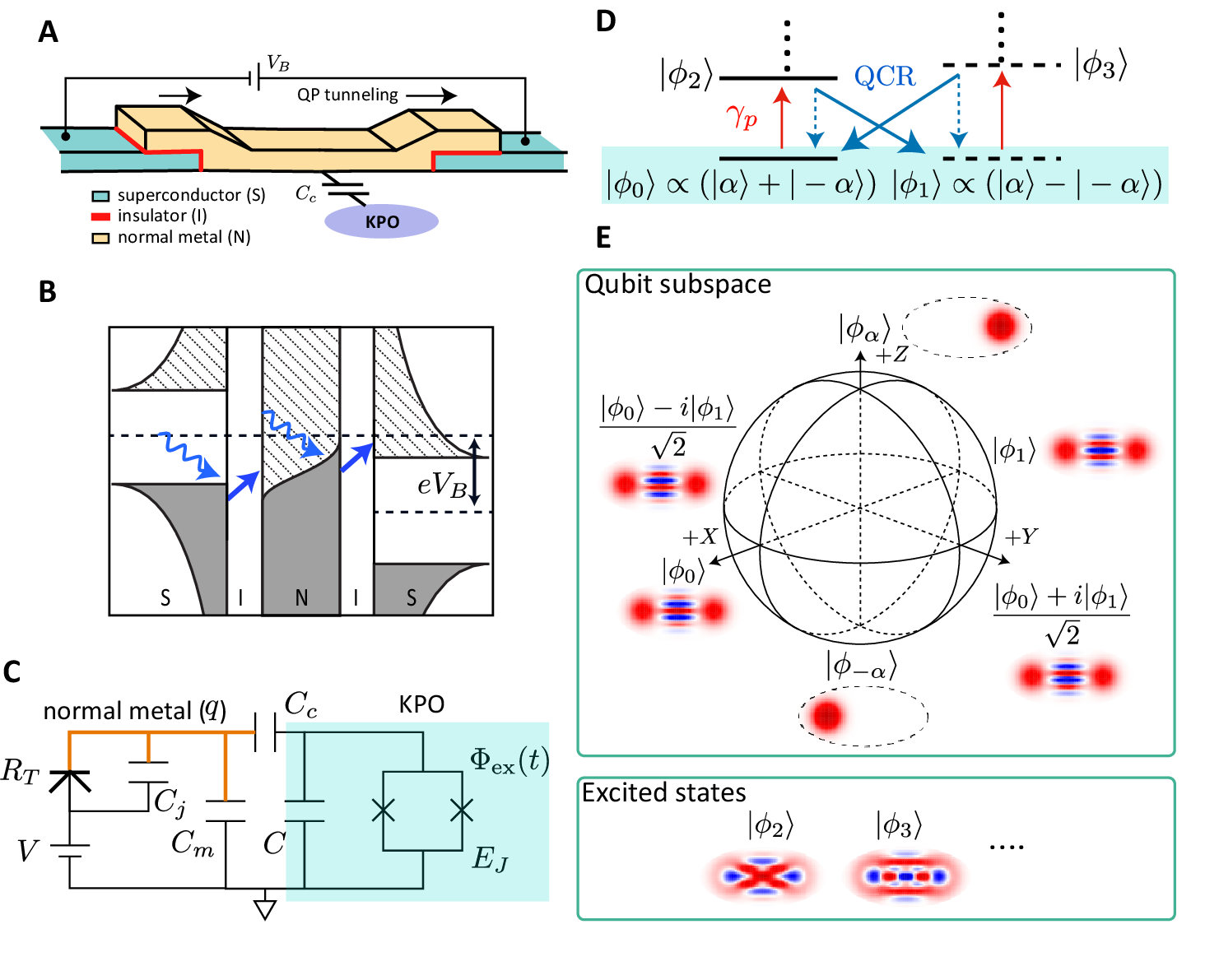}
\end{center}
\caption{
({\bf A})  Schematic of the SINIS junction coupled to the KPO.
$V_B$ is the bias voltage applied to the SINIS junction.
$C_c$ is the coupling capacitance.
The arrows indicate \textcolor{black}{quasiparticle} tunnelings.
({\bf B}) Energy diagram for the single-\textcolor{black}{quasiparticle} tunneling at a bias voltage $V_B<2\Delta/e$. 
The black solid curves at the normal metal and the superconductors represent the Fermi-Dirac distribution function and the density of states in the superconductors, respectively. The colored and shaded areas represent the occupied and unoccupied states, respectively. The straight arrows indicate the \textcolor{black}{quasiparticle} tunnelings from an initial energy state (beginning of the arrow) to a final energy state (end of the arrow). 
The wavy arrows indicate energy absorption from the KPO.
({\bf C}) Effective circuit of the system composed of a NIS junction, the KPO, and the coupling capacitance $C_c$.
The part in orange is the normal-metal island of the NIS junction with $q$ excess \textcolor{black}{quasiparticle}s, and the part shaded by light blue is the KPO formed with capacitance $C$ and a SQUID with the Josephson energy $E_J$ and an external magnetic flux $\Phi_{\rm ex}(t)$.
The circuit has only one of the NIS junctions with the junction capacitance $C_j$ and the tunneling resistance $R_T$.
$C_m$ is the capacitance of the metallic island to the ground, which includes the capacitance of another junction.
$V=V_B/2$ is the bias voltage applied to a single NIS junction.
({\bf D}) Energy diagram of the KPO. 
The horizontal solid and dashed lines represent even and odd parity energy eigenstates, respectively.
The two lowest degenerate levels shaded by light blue are qubit states.
The red arrows indicate excitations induced by the pure dephasing.
The blue solid (dashed) arrows indicate deexcitations by the QCR to opposite (the same) parity energy levels. 
({\bf E}) Bloch sphere of the Kerr-cat qubit. Schematic drawings of the Wigner functions corresponding to the eigenstates of $X$, $Y$, and $Z$ Pauli operators and excited states outside the qubit subspace are presented.
 }
\label{circuit_KPO_1_11_24}
\end{figure}

Although our theory can be applied to broader class of superconducting circuits,
we particularly consider a \textcolor{black}{Kerr parametric oscillator (KPO)} coupled to a SINIS junction of which schematic is shown in Fig.~\ref{circuit_KPO_1_11_24}(A).
When the SINIS junction is biased by a voltage $V_B$, the tunneling of quasiparticles occurs through the junctions.
The normal-metal island of the SINIS junction is capacitively coupled to the KPO.
The \textcolor{black}{quasiparticle} tunneling causes the change in the electric charge of the normal-metal island and influences the KPO via the aforementioned capacitive coupling.
The interaction between \textcolor{black}{quasiparticle}s and the KPO mediated by the normal-metal island can cause transitions between  energy levels of the KPO accompanied by the \textcolor{black}{quasiparticle} tunneling.
A tunneling \textcolor{black}{quasiparticle} can absorb energy from the KPO.
Such \textcolor{black}{quasiparticle} tunneling is sometimes called photon-assisted tunneling~\cite{Tien1963}.
The bias voltage can be used to control the rate of deexcitations (cooling) and excitations (heating) of the KPO.
 Figure~\ref{circuit_KPO_1_11_24}(B) shows the energy diagram for single-\textcolor{black}{quasiparticle} tunneling corresponding to a bias voltage where the photon-assisted \textcolor{black}{quasiparticle} tunneling is observed. 
By absorbing energy from the KPO, a \textcolor{black}{quasiparticle} can tunnel to an unoccupied higher-energy state on the opposite side of the junction.

Figure~\ref{circuit_KPO_1_11_24}(C) is the effective circuit of the system used to discuss the effect of one of the NIS junctions.
Another junction is regarded as a capacitor of which capacitance is included in the capacitance of the metallic island to the ground~\cite{Silveri2017}.
The SQUID of the KPO is subjected to an oscillating magnetic flux with the angular frequency $\omega_p$~\cite{Wang2019}.
$C_c$ is the coupling capacitance between the normal-metal island and the KPO.
The effective Hamiltonian of the KPO is written, in a rotating frame at $\omega_p/2$, as 
\begin{eqnarray}
H_{\rm KPO}^{\rm (RF)} / \hbar = \Delta_{\rm KPO} a^\dagger a - \frac{\chi}{2} a^\dagger a^\dagger a a
+\beta(a^2 + a^{\dagger 2}),
\label{HKPO_11_15_24}
\end{eqnarray}
where $\Delta_{\rm KPO}$, $\chi$, and $\beta$ are the detuning, the Kerr nonlinearity, and the amplitude of the pump field, respectively (see, e.g., Ref.~\cite{Wang2019} and 
\textcolor{black}{subsection ``Unitary transformations" in the Methods section} for the derivations and the definitions of the parameters).
In this paper, we consider the case that $\Delta_{\rm KPO}=0$.
The schematic energy diagram of the KPO is sketched in Fig.~\ref{circuit_KPO_1_11_24}(D), where
the order of the energy levels is determined by the energy in the lab frame, and is opposite to that in the rotating frame.
The two lowest energy levels are degenerate and written as
\begin{eqnarray}
|\phi_0\rangle &=& N_+ (|\alpha\rangle + |-\alpha\rangle), \nonumber\\
|\phi_1\rangle &=& N_- (|\alpha\rangle - |-\alpha\rangle), 
\end{eqnarray}
with coherent states $|\pm\alpha\rangle$, where $\alpha=\sqrt{2\beta/\chi}$ and $N_{\pm} = (2\pm 2e^{-2\alpha^2})^{-1/2}$.
These states define a Kerr-cat qubit.
Because of their degeneracy and orthogonality, $|\phi_{\pm\alpha}\rangle=(|\phi_{0}\rangle\pm |\phi_{1}\rangle)/\sqrt{2}$ are also energy eigenstates orthogonal to each other.
We have $|\phi_{\pm\alpha}\rangle \simeq |\pm\alpha\rangle$ for sufficiently large $\alpha$.
We work in a basis in which $|\phi_{\pm\alpha}\rangle$ are along the $z$-axis of the Bloch sphere [Fig.~\ref{circuit_KPO_1_11_24}(E)].
Because $H_{\rm KPO}^{\rm (RF)}$ conserves parity, its
linearly independent eigenstates can be taken so that they have either even or odd parity.
We represent energy eigenstates with even and odd parity as $|\phi_{2n}\rangle$ and $|\phi_{2n+1}\rangle$, respectively, where $n(\ge 0)$ is an integer.

As experimentally observed in Ref.~\cite{Yamaji2022} and shown numerically in 
\textcolor{black}{subsection ``Pure dephasing and single-photon loss" in the Methods section}, the pure dephasing of the KPO causes transitions from $|\phi_{0,1}\rangle$ to excited states with the same parity [Fig.~\ref{circuit_KPO_1_11_24}(D)]. 
The role of the QCR is to bring the population of the states back to the qubit subspace by absorbing excess energy from the KPO.
The main purpose of this paper is to present the cooling performance of the \textcolor{black}{quantum circuit refrigerator (QCR)} and possible drawbacks such as QCR-induced phase flip (transition between $|\phi_0\rangle$ and $|\phi_1\rangle$) and bit flip (transition between $|\phi_\alpha\rangle$ and $|\phi_{-\alpha}\rangle$).

\textcolor{black}{
\section*{RESULTS}
\subsection*{Master equation and rate of QCR-induced transitions}}
The Hamiltonian of the system composed of \textcolor{black}{quasiparticle}s in a NIS junction and the effective circuit in Fig.~\ref{circuit_KPO_1_11_24}(C) is written as
\begin{eqnarray}
H_{\rm tot} = H_{\rm QP} + H_T + H_{0},
\end{eqnarray}
where $H_{\rm QP}$ is the Hamiltonian of \textcolor{black}{quasiparticle}s given by
\begin{eqnarray}
H_{\rm QP} = \sum_{k,\sigma} (\varepsilon_{k}-eV) c_{k\sigma}^\dagger c_{k\sigma}
+ \sum_{l,\sigma} \varepsilon_{l} d_{l\sigma}^\dagger d_{l\sigma}.
\label{Hqp_1_30_23}
\end{eqnarray}
Here, $e$ is the elementary charge\textcolor{black}{, and subscript QP indicates quasiparticle}. $c_{k\sigma}$ and $d_{l\sigma}$ are the annihilation operators for \textcolor{black}{quasiparticle}s in the superconducting electrode and the normal-metal island, respectively.
$\varepsilon_{k}$ and $\varepsilon_{l}$ are the energies of \textcolor{black}{quasiparticle}s with wave numbers $k$ and $l$, while $\sigma$ denotes their spins. 
In our model, \textcolor{black}{quasiparticle}s in the superconducting electrodes are treated as \textcolor{black}{quasiparticle}s in the normal-metal island except that the density of states is different (the semiconductor model)~\cite{Ingold1992,Tinkham_textbook}.
The energy shift of $-eV$ in the first term represents the effect of the bias voltage $V=V_B/2$. 
Tunneling Hamiltonian $H_T$ represents the tunneling of \textcolor{black}{quasiparticle}s and the interaction between \textcolor{black}{quasiparticle}s and the superconducting circuit, and is written as~\cite{Silveri2017}
\begin{eqnarray}
H_T = \sum_{k,l,\sigma} T_{lk} d_{l\sigma}^\dagger c_{k\sigma} e^{-i\varphi_N} + h.c.,
\label{HT_8_1_23}
\end{eqnarray}
where $\varphi_N$ is a dimensionless flux, that is, $\hbar \varphi_N/e$ is the flux, $\Phi_N$, defined by time integration of the node voltage.
The factor $e^{-i\varphi_N}$ represents the shift of the electric charge in the normal-metal island accompanied by a \textcolor{black}{quasiparticle} tunneling.
The Hamiltonian of the effective circuit is written as
\begin{eqnarray}
H_0 = \frac{(Q_N + Q_j)^2}{2C_N} + \frac{[Q+\alpha_{c}(Q_N + Q_j)]^2}{2C_r} - E_J(t) \cos\Big{(}\frac{2e}{\hbar}\Phi\Big{)},
\end{eqnarray} 
where $C_N = C_c + C_m + C_j$, $C_r = C + \alpha_{c} {C_\Sigma}_m$, $\alpha_{c} =  C_c/C_N$,
${C_\Sigma}_m = C_m + C_j$, $Q_j = C_j V$, and $Q_N$ is the conjugate charge of the flux at the normal-metal island $\Phi_N$ (see~\cite{Silveri2017} for the derivation of the Hamiltonian of a similar circuit).
$\Phi$ is the flux at the \textcolor{black}{Kerr parametric oscillator (KPO)} and $Q$ is its conjugate charge.
We have the commutation relations, $[\Phi,Q]=[\Phi_N,Q_N]=i\hbar$.
We assume that the Josephson energy $E_J(t)$ is modulated  as $E_J(t)=E_J+\delta E_J \cos(\omega_p t)$ via a time-dependent magnetic flux in the SQUID \textcolor{black}{(see subsection ``Unitary transformations" in the Methods section)}.

For later convenience and for moving into the rotating frame at frequency $\omega_{\rm RF}=\omega_p/2$, we apply unitary transformations, by which ${H}_T$ and ${H}_{0}$ are transformed to ${H}_T^{\rm (RF)}$ and ${H}_{0}^{\rm (RF)}$ while $H_{\rm QP}$ is unchanged (see \textcolor{black}{subsection ``Unitary transformations" in the Methods section} for details). 
In the rotating wave approximation, 
${H}_{0}^{\rm (RF)}$ is written as
\begin{eqnarray} 
{H}_{0}^{\rm (RF)}  = \sum_q \Big{[} \frac{e^2q^2}{2C_N}|q\rangle \langle q|  \Big{]}+ H_{\rm KPO}^{\rm (RF)} ,
\label{H0_RF_12_19_23}
\end{eqnarray} 
where $Q_N|q\rangle=eq|q\rangle$, and $q$ is an integer denoting the excess charge number in the normal-metal island.
On the other hand, $H_T^{\rm (RF)}$ can be written as 
\begin{eqnarray}
H_T^{\rm (RF)}(t) &=& \sum_{m,\delta m,q} \sum_{k,l,\sigma} e^{i\omega_{\rm RF} \delta m t} \Big\{
\langle \delta m+m|  \exp\Big{[} -\frac{i}{\hbar} \alpha_{c} e \Phi \Big{]}  |m\rangle T_{lk} d_{l\sigma}^\dagger c_{k\sigma} |q-1,\delta m+m\rangle \langle q,m| \nonumber\\
&& + \langle \delta m +m|  \exp\Big{[} \frac{i}{\hbar} \alpha_{c} e \Phi \Big{]}  |m\rangle
T_{lk}^\ast c_{k\sigma}^\dagger  d_{l\sigma} |q+1,\delta m + m\rangle \langle q,m| \Big\},
\label{HRF_2_12_24}
\end{eqnarray}
where $|m\rangle$ denotes a Fock state of the KPO.
The first and second indices of $|q,m\rangle (= |q\rangle \otimes  |m\rangle)$ denote the state of the normal-metal island and the KPO, respectively.
We will regard $H_T^{\rm (RF)}(t)$ as a perturbation in the derivation of the master equation for the KPO.

Suppose that $|\psi_{\mu(\mu')}\rangle$ is an eigenstate of $H=H_{\rm QP} + H_{0}^{\rm (RF)}$ with energy $E_{\mu (\mu')}$, and we have 
\begin{eqnarray} 
H |\psi_\mu\rangle = E_\mu |\psi_\mu\rangle, \hspace{0.5cm} \langle \psi_\mu | \psi_{\mu'}\rangle=\delta_{\mu{\mu'}}.
\end{eqnarray}
We divide the system into two parts: the KPO and the others, that is, \textcolor{black}{quasiparticle}s and the normal-metal island.
The latter is called environment. 
An eigenstate of $H$ can be expressed as $|\psi_\mu\rangle=|\phi_{\mu}, \mathcal{E}_\mu\rangle (= |\phi_{\mu}\rangle \otimes |\mathcal{E}_\mu\rangle)$,
where $|\phi_{\mu}\rangle$ is an eigenstate of $H_{\rm KPO}^{\rm (RF)}$ with energy $E_{\phi_{\mu}}$ and $|\mathcal{E}_\mu\rangle$ denotes a state of the environment with energy $E_{\mathcal{E}_{\mu}}$.
The reduced density matrix for the KPO is obtained by tracing out the environment, and is written
 as \begin{eqnarray}
\rho_{\rm KPO}(t) = \sum_{\mathcal{E}_\mu} \langle \mathcal{E}_\mu |  \rho(t) |  \mathcal{E}_\mu \rangle = \sum_{\phi_{\mu}, \phi_{\mu'}} \rho^{\rm KPO}_{\phi_{\mu}, \phi_{\mu'}}(t) |\phi_{\mu}\rangle \langle \phi_{\mu'}|
\label{rho_KPO_12_11_24}
\end{eqnarray} 
with \textcolor{black}{the density matrix of the total system $\rho(t)$ and} the matrix elements given by
\begin{eqnarray}
\rho^{\rm KPO}_{\phi_{\mu}, \phi_{\mu'}}(t) 
= \sum_{\mathcal{E}_\mu} \langle \phi_{\mu}, \mathcal{E}_\mu | \rho(t) | \phi_{\mu'}, \mathcal{E}_\mu \rangle.
\label{rho_sys_2_14_24}
\end{eqnarray} 

We can derive the master equation for the KPO by tracing out the environment from the equation of motion for the total system and also by taking into account the effect of another NIS junction (see \textcolor{black}{subsection ``Derivation of master equation" in}
the Methods section for the derivation).
The master equation is written as
\begin{eqnarray}
\rho_{\phi_{\mu},\phi_{\mu'}}^{\rm KPO}(t+\Delta t) &=& \rho_{\phi_{\mu},\phi_{\mu'}}^{\rm KPO}(t)
- i\omega_{\phi_{\mu},\phi_{\mu'}} \Delta t \rho_{\phi_{\mu},\phi_{\mu'}}^{\rm KPO}(t) \nonumber\\
&& + \sum_{\phi_\nu}\sum_{\phi_{\nu'}}' \Gamma^{(1)}(\phi_{\mu},\phi_{\mu'},\phi_\nu,\phi_{\nu'},V) \Delta t
\rho_{\phi_\nu,\phi_{\nu'}}^{\rm KPO}(t) \nonumber\\
&& + \sum_{\phi_{\xi}}' \Gamma^{(2)}(\phi_{\mu},\phi_{\mu'},\phi_{\xi},V) \Delta t
\rho_{\phi_\xi,\phi_{\mu'}}^{\rm KPO}(t)\nonumber\\
&& + \sum_{\phi_{\xi}}'' \Gamma^{(3)}(\phi_{\mu},\phi_{\mu'},\phi_{\xi},V) \Delta t
\rho_{\phi_{\mu},\phi_{\xi}}^{\rm KPO}(t),
\label{EOM_10_24_23}
\end{eqnarray} 
where $\omega_{\phi_{\mu},\phi_{\mu'}} =(E_{\phi_{\mu}} - E_{\phi_{\mu'}})/\hbar$, and
\begin{eqnarray}
\Gamma^{(1)}(\phi_{\mu},\phi_{\mu'},\phi_\nu,\phi_{\nu'},V) &=&
\frac{2}{e^2{R_T}} \sum_{\delta m,\delta m',q}  p_{q}  \nonumber\\
&& \times \Big{[} \int d\varepsilon_k n_s(\varepsilon_k)
[1-f(\varepsilon_k,T_S)] f(\varepsilon_l^{(f,\delta m,1)},T_N)
\eta_{\phi_{\mu},\phi_{\nu}}^{(f,\delta m)} \big{(}\eta_{\phi_{\mu'},\phi_{\nu'}}^{(f,\delta m')}\big{)}^\ast 
\nonumber\\
&& + \int d\varepsilon_k n_s(\varepsilon_k)
f(\varepsilon_k,T_S) [1-f(\varepsilon_l^{(b,\delta m,1)},T_N)]
\eta_{\phi_{\mu},\phi_{\nu}}^{(b,\delta m)} (\eta_{\phi_{\mu'},\phi_{\nu'}}^{(b,\delta m')})^\ast \Big{]},
\nonumber\\
\Gamma^{(2)}(\phi_{\mu},\phi_{\mu'},\phi_\xi,V) &=&
-\frac{1}{e^2{R_T}} \sum_{\delta m,\delta m',q} \sum_{\phi_\nu} p_{q}  \nonumber\\
&& \times \Big{[} \int d\varepsilon_k n_s(\varepsilon_k)
[1-f(\varepsilon_k,T_S)] f(\varepsilon_l^{(f,\delta m, 2)},T_N)
\big{(}\eta_{\phi_{\nu},\phi_{\mu}}^{(f,\delta m)}\big{)}^\ast \eta_{\phi_{\nu},\phi_{\xi}}^{(f,\delta m')} 
\nonumber\\
&& + \int d\varepsilon_k n_s(\varepsilon_k)
f(\varepsilon_k,T_S) [1-f(\varepsilon_l^{(b,\delta m,2)},T_N)]
\big{(} \eta_{\phi_{\nu},\phi_{\mu}}^{(b,\delta m)} \big{)}^\ast \eta_{\phi_{\nu},\phi_{\xi}}^{(b,\delta m')} \Big{]},
 \nonumber\\
\Gamma^{(3)}(\phi_{\mu},\phi_{\mu'},\phi_\xi,V) &=&
-\frac{1}{e^2{R_T}} \sum_{\delta m,\delta m',q} \sum_{\phi_\nu} p_{q}  \nonumber\\
&& \times \Big{[} \int d\varepsilon_k n_s(\varepsilon_k)
[1-f(\varepsilon_k,T_S)] f(\varepsilon_l^{(f,\delta m,3)},T_N)
\eta_{\phi_\nu,\phi_{\mu'}}^{(f,\delta m)} \big{(}\eta_{\phi_{\nu},\phi_{\xi}}^{(f,\delta m')}\big{)}^\ast  
\nonumber\\
&& + \int d\varepsilon_k n_s(\varepsilon_k)
f(\varepsilon_k,T_S) [1-f(\varepsilon_l^{(b,\delta m,3)},T_N)]
\eta_{\phi_\nu,\phi_{\mu'}}^{(b,\delta m)} \big{(}\eta_{\phi_{\nu},\phi_{\xi}}^{(b,\delta m')}\big{)}^\ast  \Big{]},
 \nonumber\\
 \label{Gamma_12_21_23}
\end{eqnarray} 
and $n_s$ is the density of states of the \textcolor{black}{quasiparticle}s in the superconducting electrode given by
\begin{eqnarray}
n_s(\varepsilon) = \Big{|}{\rm Re} \Big\{ \frac{\varepsilon + i\gamma_{\rm D}\Delta}{\sqrt{(\varepsilon + i\gamma_{\rm D}\Delta)^2 - \Delta^2}} \Big\} \Big{|},
\end{eqnarray} 
with the superconductor gap parameter $\Delta$ and the Dynes parameter $\gamma_{\rm D}$~\cite{Dynes1978}.
The Fermi-Dirac distribution function is defined by $f(E,T) = 1/[e^{E/(k_B T)}+1]$ with $k_B$ the Boltzmann constant, and $T_N$ and $T_S$ the electron temperature at the normal-metal island and the superconducting electrodes, respectively.
The probability, denoted by $p_q$, that the state of the normal-metal island is $|q\rangle$, is determined using the elastic tunneling of \textcolor{black}{quasiparticle}s, in which \textcolor{black}{quasiparticle}s do not exchange energy with the KPO (see \textcolor{black}{subsection ``Probability $p_q$" in the Methods section}). 
In Eq.~(\ref{Gamma_12_21_23}), $\sum_{\phi_{\nu'}}'$, $\sum_{\phi_\xi}'$, and $\sum_{\phi_\xi}''$, respectively, denote the summation with respect to the state of the KPO, $\phi_{\nu'}$, $\phi_{\xi}$, and $\phi_{\xi}$, which satisfy
\begin{eqnarray}
E_{\phi_{\mu}} - E_{\phi_{\nu}} + \frac{\hbar\omega_p\delta m}{2} &=& E_{\phi_{\mu'}} - E_{\phi_{\nu'}} + \frac{\hbar\omega_p\delta m'}{2}, \nonumber\\
-E_{\phi_{\mu}}  + \frac{\hbar\omega_p\delta m}{2} &=& -E_{\phi_{\xi}} + \frac{\hbar\omega_p\delta m'}{2},\nonumber\\
-E_{\phi_{\mu'}}  + \frac{\hbar\omega_p\delta m}{2} &=& -E_{\phi_{\xi}} + \frac{\hbar\omega_p\delta m'}{2}.
\label{matching_12_21_23}
\end{eqnarray} 
And, $\eta_{\phi_{\mu},\phi_{\nu}}^{(f,\delta m)}$ and $\eta_{\phi_{\mu},\phi_{\nu}}^{(b,\delta m)}$ are defined by
\begin{eqnarray} 
\eta_{\phi_{\mu},\phi_{\nu}}^{(f,\delta m)} &=& \sum_m \langle \delta m + m | D(i\rho_c^{\frac{1}{2}}) | m\rangle \langle \phi_{\mu} | \delta m + m\rangle 
\langle m| \phi_\nu\rangle\nonumber\\
\eta_{\phi_{\mu},\phi_{\nu}}^{(b,\delta m)} &=& \sum_m \langle \delta m + m | D(-i\rho_c^{\frac{1}{2}}) | m\rangle \langle \phi_{\mu} | \delta m + m\rangle 
\langle m| \phi_\nu\rangle\nonumber\\
&=& (\eta_{\phi_\nu,\phi_{\mu}}^{(f,-\delta m)})^\ast,
\label{eta_5_26_24}
\end{eqnarray} 
while $\varepsilon_l^{(f,\delta m,i)}$ and $\varepsilon_l^{(b,\delta m,i)}$ for $i=1,2,3$ are defined by
\begin{eqnarray}
\varepsilon_l^{(f,\delta m,1)}  &=& 
E_{\phi_{\mu}}  - E_{\phi_\nu} + \varepsilon_k - eV + E_N(1+2q)
+\frac{\hbar \omega_p \delta m}{2},
\nonumber\\
\varepsilon_l^{(f,\delta m,2)}  &=& 
E_{\phi_{\nu}} - E_{\phi_{\mu}} + \varepsilon_k - eV + E_N(1+2q) + \frac{\hbar\omega_p\delta m}{2},
\nonumber\\
\varepsilon_l^{(f,\delta m,3)}  &=& 
E_{\phi_{\nu}} - E_{\phi_{\mu'}} + \varepsilon_k - eV + E_N(1+2q) + \frac{\hbar\omega_p\delta m}{2},
\nonumber\\
\varepsilon_l^{(b,\delta m,1)}  &=& 
E_{\phi_{\nu}} - E_{\phi_{\mu}} + \varepsilon_k - eV - E_N(1-2q) - \frac{\hbar\omega_p\delta m}{2},
\nonumber\\
\varepsilon_l^{(b,\delta m,2)}  &=&
E_{\phi_{\mu}} - E_{\phi_{\nu}} + \varepsilon_k - eV - E_N(1-2q) - \frac{\hbar\omega_p\delta m}{2},
\nonumber\\
\varepsilon_l^{(b,\delta m,3)}  &=& 
E_{\phi_{\mu'}} - E_{\phi_{\nu}} + \varepsilon_k - eV - E_N(1-2q) - \frac{\hbar\omega_p\delta m}{2}.
\end{eqnarray} 
\textcolor{black}{Here, $E_N=e^2/(2C_N)$.} 
In Eq.~(\ref{eta_5_26_24}), $\rho_c$ is the interaction parameter defined by $\rho_c=\alpha_c^2 \sqrt{E_C/ (8E_J)} $ with $E_C=e^2/(2C_r)$.
The translation operator $D(X)$ is defined as $D(X)=\exp[X a^\dagger - X^\ast a]$.
$\langle m' | D( i\rho_c^{\frac{1}{2}}) | m\rangle$ is given as \cite{Hollenhorst1979}
\begin{eqnarray}
\langle m' | D(i\rho_c^{\frac{1}{2}}) | m\rangle = 
  \left\{
    \begin{array}{l}
      e^{-\frac{\rho_c}{2}} i^l \rho_c^{\frac{l}{2}} \sqrt{\frac{m'!}{m!}} L_{m'}^l(\rho_c)\ \ \ \ \ {\rm for} \ \ m\ge m', \\
      e^{-\frac{\rho_c}{2}} i^{-l} \rho_c^{-\frac{l}{2}} \sqrt{\frac{m!}{m'!}} L_{m}^{-l}(\rho_c)\ \ {\rm for} \ \ m< m',
    \end{array}
  \right.
\end{eqnarray}
where $l=m-m'$, and $L_{m}^l$ is the generalized Laguerre polynomials.
The superscript $f$ of $\eta_{\phi_{\mu},\phi_{\nu}}$ and $\varepsilon_l$ denotes the forward tunneling where the number of \textcolor{black}{quasiparticle}s increases in the normal-metal island, while $b$ denotes opposite (backward) tunneling.
Because of Eq.~(\ref{EOM_10_24_23}), $\Gamma^{(1)}(\phi_{i},\phi_{i},\phi_{j},\phi_{j},V)$ can be regarded as the rate of the transition from $|\phi_{j}\rangle$ to $|\phi_{i}\rangle$ caused by the QCR, and is quantitatively studied in the following section. The other $\Gamma$s are also important to describe the dynamics of the KPO.

\subsection*{Cooling performance of \textcolor{black}{quantum circuit refrigerator (QCR)}}
\begin{figure}
\begin{center}
\includegraphics[width=15cm]{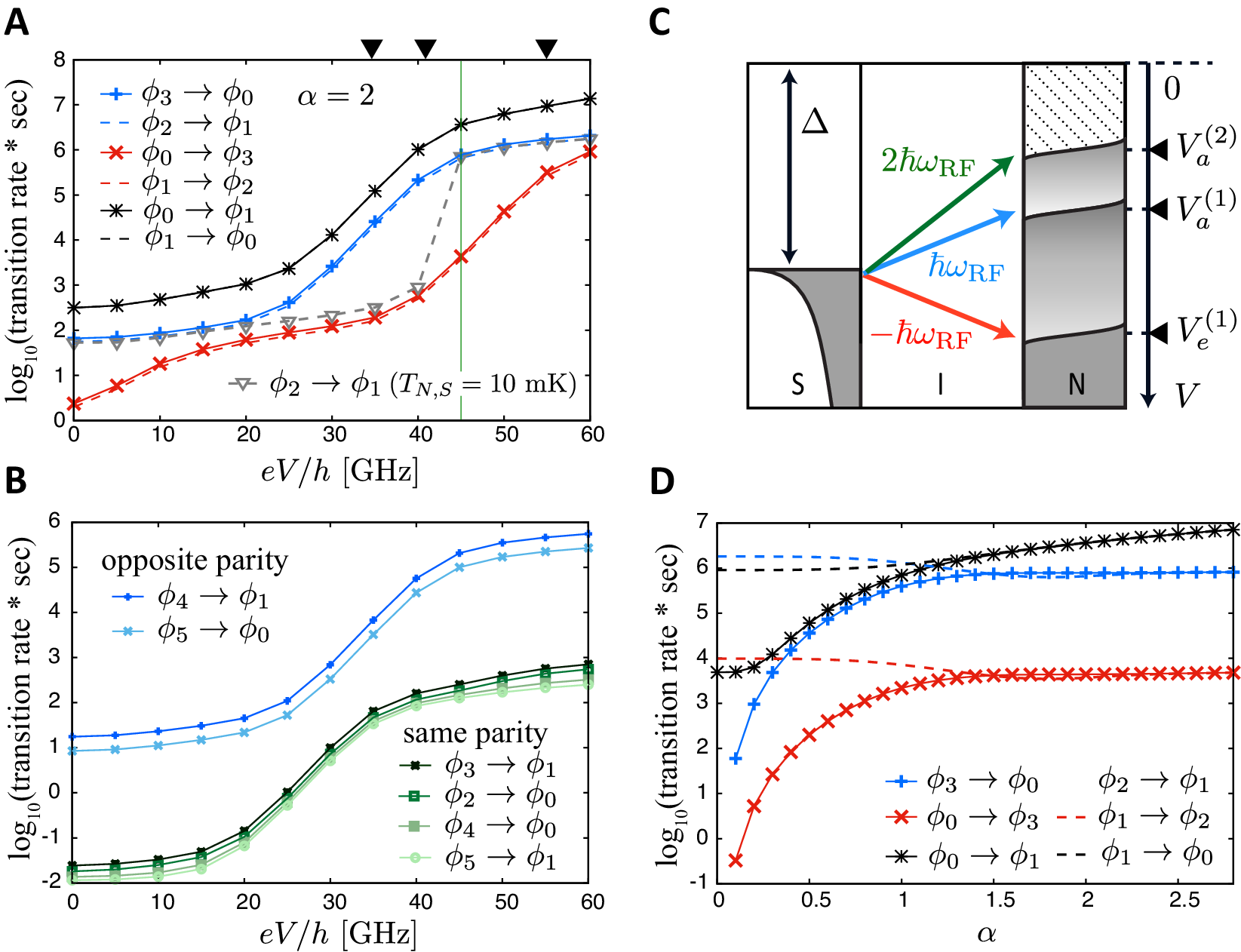}
\end{center}
\caption{
({\bf A}) Voltage dependence of the rates of dominant transitions corresponding to phase flip (black), cooling (blue), and heating (red) of the KPO for $\alpha=2$. 
The parameters \textcolor{black}{used} are \textcolor{black}{$\rho_c=5\times 10^{-5}$}, $\chi/2\pi=10$~MHz, $\omega_c/2\pi=$7~GHz, $\Delta=200$~$\mu$eV,
$\Delta_{\rm KPO}/2\pi=0$~MHz, $R_{\rm T}=50$~k$\Omega$, $\gamma_{\rm D}=10^{-4}$, $\beta/2\pi=20$~MHz, and $T_{N,S}=100$~mK.
The values of the parameters\textcolor{black}{, except for $\rho_c$,} are comparable to the ones measured or used in the experiments~\cite{Yamaji2022,Tan2017}. 
\textcolor{black}{The value of $\rho_c$ is smaller than that used in the previous work~\cite{Silveri2017}.}
The data points in grey color are for $T_{N,S}=10~$mK.
The voltage indicated by the black triangles are $V^{(2)}_a$, $V^{(1)}_a$, and $V^{(1)}_e$ in panel (C).
The green vertical line indicates $eV/h=45$~GHz used in panel (D).
({\bf B}) The same things as panel (A) but for other transitions from excited states to the qubit states.
({\bf C}) Schematic of energy diagram of a NIS junction. The dark green, light blue and red arrows indicate two-photon-absorption, single-photon-absorption, and single-photon-emission processes, respectively. The minimum voltages at which these processes can occur are $V^{(2)}_a$, $V^{(1)}_a$, and $V^{(1)}_e$ for $T_{N,S}=0$, respectively. 
We have $eV^{(2)}_a/h\simeq 34$~GHz, $eV^{(1)}_a/h\simeq $41~GHz, and $eV^{(1)}_e/h\simeq $55~GHz for the  parameters \textcolor{black}{used}.
({\bf D}) $\alpha$ dependence of relevant transition rates for $eV/h=45$~GHz. 
The color scheme is the same as in panel (A).}
\label{Gamma1_com2_12_11_23}
\end{figure}

We quantitatively examine the cooling performance of the QCR which is controlled with the bias voltage $V$.
\textcolor{black}{The results presented in the figures below are obtained through numerical simulations. }
Figure~\ref{Gamma1_com2_12_11_23}(A) shows the voltage dependence of the dominant transition rates relevant to cooling, heating, and phase flip of the \textcolor{black}{Kerr parametric oscillator (KPO)} for experimentally feasible parameters, while Fig.~\ref{Gamma1_com2_12_11_23}(B) shows the rates of other transitions to the qubit states.
The transition rates from excited states to the qubit states (cooling rates) can be changed by more than four orders of magnitude for the used parameters.
The cooling rates dominate over heating rates, especially for \textcolor{black}{30}~GHz $<eV/h<50$~GHz as shown in Fig.~\ref{Gamma1_com2_12_11_23}(A).
These results suggest that the QCR can serve as an on-chip refrigerator for the KPO reducing the population of excited states, and the cooling power can be tuned over a wide range by the bias voltage.
The phase-flip rate also increases with $V$ as do the cooling rates. 
\textcolor{black}{We note that the results in Figs.~\ref{Gamma1_com2_12_11_23}(A), \ref{Gamma1_com2_12_11_23}(B), and \ref{Gamma1_com2_12_11_23}(D) are the rates of the QCR-induced transitions.
In our theory, these rates are independent of the transitions caused by other decoherence sources such as the single-photon loss and the pure dephasing, which are considered later.}

The bias voltage dependence of the transition rates can be understood from an energy diagram of \textcolor{black}{a normal-metal--insulator--superconductor (NIS)} junction for single-\textcolor{black}{quasiparticle} tunnelings [Fig.~\ref{Gamma1_com2_12_11_23}(C)], which also shows the minimum voltage at which each type of photon-assisted electron tunnelings can occur for $T_{N,S}=0$.
The voltage is given by $V_a^{(2)}=(\Delta-2\hbar\omega_{\rm RF})/e$, $V_a^{(1)}=(\Delta-\hbar\omega_{\rm RF})/e$, and $V_e^{(1)}=(\Delta+\hbar\omega_{\rm RF})/e$ for two-photon-absorption, single-photon-absorption, and single-photon-emission processes, respectively. 
The rate for transition $|\phi_{2}\rangle \rightarrow |\phi_1\rangle$ jumps at around $V=V_a^{(1)}$ for sufficiently low $T_{N,S}$ as seen in the result at $T_{N,S}=10$~mK in Fig.~\ref{Gamma1_com2_12_11_23}(A).
It suggests that this transition is due to the single-photon-absorption process. 
For the same reason, we consider that the opposite-parity (same-parity) transitions in Fig.~\ref{Gamma1_com2_12_11_23}(A,B) are caused mainly by a single-photon (two-photon) process. 
Note that two-photon-absorption processes can occur at a smaller voltage than the single-photon-absorption processes.
On the other hand, with more experimentally feasible temperatures ($T_{N,S}=100$~mK), the increase of transition rates with respect to $V$ is gradual and starts at smaller voltages than that for $T_{N,S}=0$.
This is due to the temperature effect of the normal-metal island, which has a smooth variation of the Fermi distribution function.
Figure~\ref{Gamma1_com2_12_11_23}(D) shows the $\alpha$ dependence of relevant transition rates for $eV/h=45$~GHz.
It is seen that the cooling and heating rates are insensitive to $\alpha$ for $\alpha>1.5$ which is in the typical parameter regime of the Kerr-cat qubit, and the cooling rates are two-orders-of-magnitude higher than the heating rates.
Therefore, the cooling effect is robust against changes in $\alpha$. 
On the other hand, the phase-flip rate monotonically increases with $\alpha$. 
We attribute this to the facts that the QCR approximately works as a source of single-photon loss with this bias voltage regime, and the single-photon loss causes the effective phase flip with the rate proportional to $|\alpha|^2$~\cite{Puri2017b}.
We also note that $\alpha=0$ corresponds to a trasmon-type qubit, and the rate of the transition from $|\phi_1\rangle$ to $|\phi_0\rangle$ is the cooling rate for the transmon, while the rate of the transition from $|\phi_0\rangle$ to $|\phi_1\rangle$ is the heating rate. 




\subsection*{Dynamics and stationary state of a KPO under operation of QCR}
\begin{figure}
\begin{center}
\includegraphics[width=15cm]{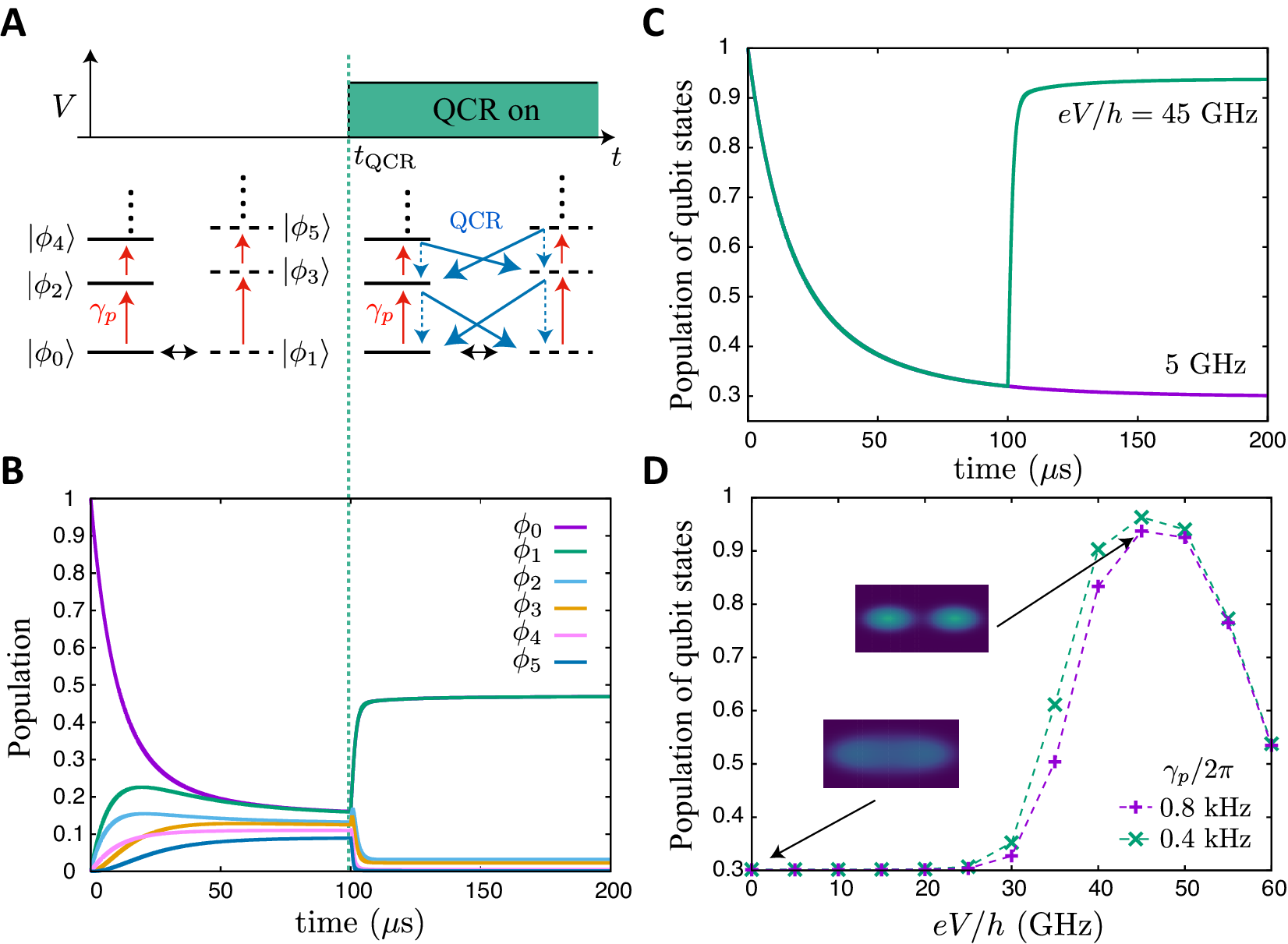}
\end{center}
\caption{
({\bf A}) Schedule of the QCR and relevant inter-level transitions when the QCR is off (left) and on (right).
Time dependence of the population of energy levels $P_i$~({\bf B}) and the population of the qubit states $P_0+P_1$~({\bf C}) for \textcolor{black}{$\kappa/2\pi=1.6$~kHz} and \textcolor{black}{$\gamma_p/2\pi=0.8$~kHz}. 
The values of $\kappa$ and $\gamma_p$ are comparable to the ones measured \textcolor{black}{for flux-tunable superconducting qubits}~\cite{Chavez-Garcia2022,Tuokkola2024}.
({\bf D}) Population of the qubit states of the stationary state as a function of $V$ for $\kappa=2\gamma_p$.
Insets are the Husimi Q function of the stationary states for \textcolor{black}{$\gamma_p/2\pi=0.8$~kHz}.
The other parameters used are the same as in Fig.~\ref{Gamma1_com2_12_11_23}.}
\label{population_2_5_24}
\end{figure}

We study dynamics and stationary states of the KPO under operation of the QCR.
We assume that initially the QCR is off and the state of the KPO is $|\phi_0\rangle$. 
The QCR is turned on at $t=t_{\rm QCR}$ as illustrated in Fig.~\ref{population_2_5_24}(A).
We take into account the pure dephasing $\gamma_p$ and the single-photon loss $\kappa$ of the KPO which are not derived from the QCR, by including the second and the third terms of Eq.~(\ref{ME_wo_QCR_4_8_24}) in our equation of motion in addition to the effect of the QCR represented by Eq.~(\ref{EOM_10_24_23}) \textcolor{black}{(see subsection ``Pure dephasing and single-photon loss" in the Methods section)}.
To distinguish the single-photon loss from the QCR-induced photon loss, we refer to it as original single-photon loss.
Relevant inter-level transitions for $t<t_{\rm QCR}$ and $t> t_{\rm QCR}$ are illustrated in Fig.~\ref{population_2_5_24}(A).

Figure~\ref{population_2_5_24}(B) shows the time dependence of $P_i$, the population of $|\phi_{i}\rangle$.
For $t<t_{\rm QCR}$, $P_{i(>0)}$ increases with time while $P_{0}$ decreases.
\textcolor{black}{(We numerically integrated the equation of motion using a fourth order Runge-Kutta method in order to simulate the dynamics of the KPO.)}
The increase of $P_1$ is due to the phase flip from $|\phi_{0}\rangle$ to $|\phi_{1}\rangle$, indicated by a black arrow in Fig.~\ref{population_2_5_24}(A), caused by the original single-photon loss. 
The increase of $P_{i(>1)}$ is due to heating induced by the pure dephasing denoted by the red arrows.
For $t>t_{\rm QCR}$, $P_{i(>1)}$ decreases and $P_{i(\le 1)}$ increases due to the cooling effect represented by blue arrows.
For sufficiently large $t$,  $|\phi_{0}\rangle$ and $|\phi_{1}\rangle$ are equally populated due to the phase flip.
The population of the qubit states $P_0+P_1$ increases at $t=t_{\rm QCR}$ for $eV/h=45$~GHz as shown in Fig.~\ref{population_2_5_24}(C), because of high QCR-induced cooling rates dominating over $\gamma_p$.
An increase in $P_0+P_1$ is not seen for $eV/h=5$~GHz because $\kappa$ and $\gamma_p$ govern the dynamics of the KPO.
In Fig.~\ref{population_2_5_24}(D) the population of the qubit states for the stationary state is exhibited as a function of the bias voltage for two different $\gamma_p$ with $\kappa=2\gamma_p$.
The population can be higher than \textcolor{black}{0.93} by tuning the bias voltage while it is about \textcolor{black}{0.3} when the QCR is off. 
The population for smaller $\gamma_p$ becomes higher than for larger $\gamma_p$ at $eV/h=35$~GHz.
This is because less cooling rate is sufficient to see the effectiveness of the QCR for smaller $\gamma_p$.
In the small voltage regime $eV/h\le 15$~GHz, the QCR is effectively off, and the population is determined by the ratio $\gamma_p/\kappa$.
On the other hand, for large voltage regime $eV/h\ge 55$~GHz, QCR-induced transitions dominate the effect of $\gamma_p$ and $\kappa$. Therefore, the population is insensitive to $\gamma_p$ amd $\kappa$.

\subsection*{Biased nature of noise is preserved under operation of QCR}
So far we studied the effect of the QCR especially on the population $P_i$, which is the diagonal element of the density matrix of the KPO $\rho^{\rm KPO}_{\phi_i,\phi_i}$.
Now we study the effect of the QCR on the off-diagonal elements of the density matrix, and present that the biased nature of errors of the KPO is preserved even under operation of the QCR, that is, the bit-flip rate is much smaller than the phase-flip rate. 
In order to see the impact of the QCR on the off-diagonal elements, we assume that the initial state is $|\phi_\alpha\rangle\propto |\phi_0\rangle + |\phi_1\rangle$.
If decoherence caused by the QCR significantly enhances the decay of the off-diagonal elements, $\rho^{\rm KPO}_{\phi_0,\phi_1}$ and $\rho^{\rm KPO}_{\phi_1,\phi_0}$, the KPO rapidly approaches the mixed state of $|\phi_\alpha\rangle$ and $|\phi_{-\alpha}\rangle$. This can be interpreted as the QCR enhances bit flips, and the biased nature of errors of the KPO is lost.
Importantly, as shown below, such decoherence is suppressed when $|\phi_0\rangle$ and $|\phi_1\rangle$ are degenerate.

\begin{figure}
\begin{center}
\includegraphics[width=14cm]{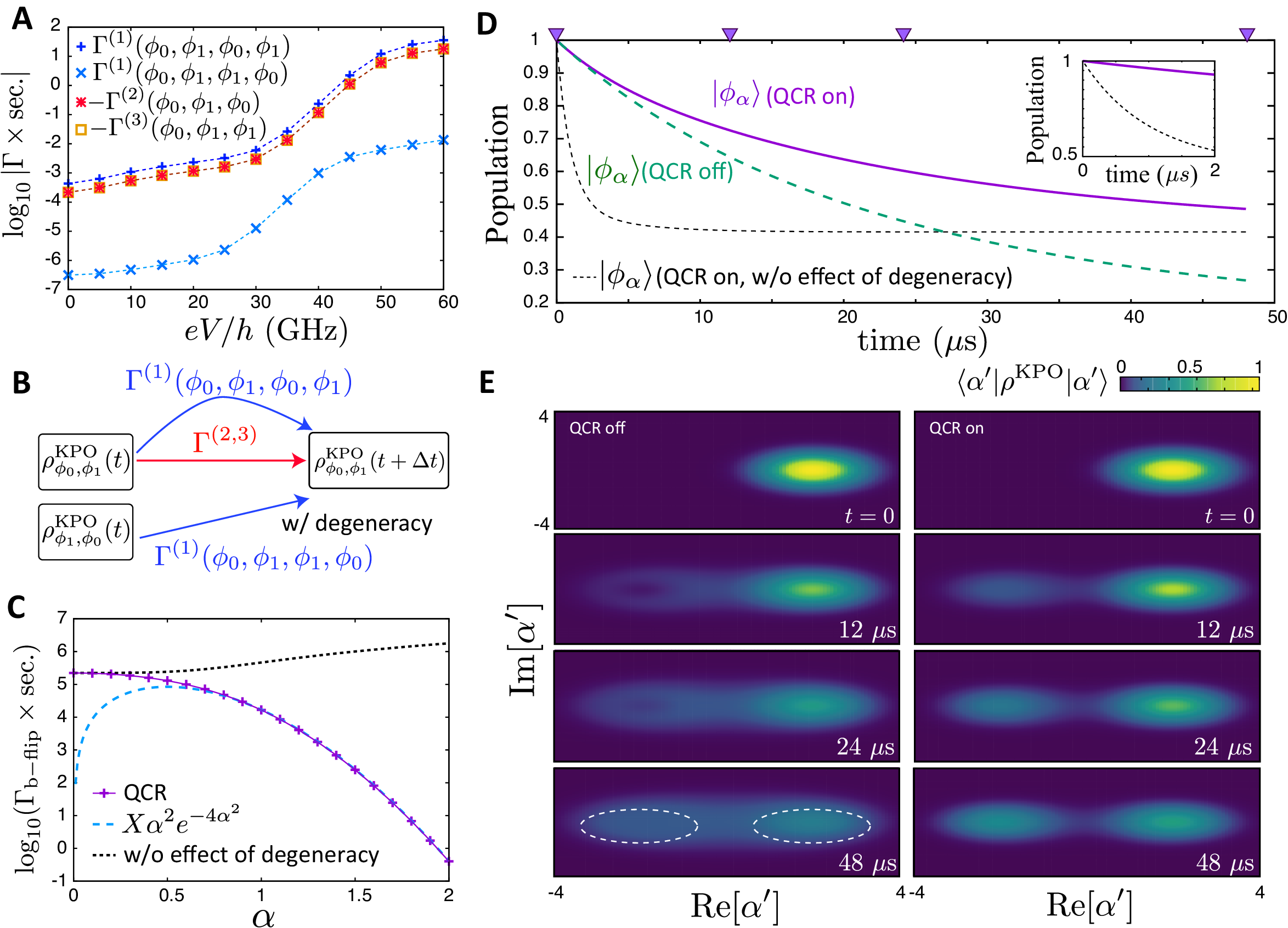}
\end{center}
\caption{
({\bf A}) $\Gamma$s relevant to the change in $\rho^{\rm KPO}_{\phi_0,\phi_1}$ as a function of the voltage $V$ applied to a NIS junction.
The used parameters are the same as in Fig.~\ref{Gamma1_com2_12_11_23}.
({\bf B}) Schematic of the role of the relevant $\Gamma$s for the change of $\rho^{\rm KPO}_{\phi_0,\phi_1}$.
$\Gamma^{(2)}$ and $\Gamma^{(3)}$ abbreviate $\Gamma^{(2)}(\phi_0,\phi_1,\phi_0)$ and $\Gamma^{(3)}(\phi_0,\phi_1,\phi_1)$, respectively.
The blue (red) arrows indicate that positive (negative) $\Gamma$s preserve (degrade) the coherence of the KPO. $\Gamma^{(1)}(\phi_0,\phi_1,\phi_1,\phi_0)$ is the effect of the quantum interference arising from the degeneracy between $|\phi_0\rangle$ and $|\phi_1\rangle$.
({\bf C}) The QCR-induced bit-flip rate as a function of $\alpha$. 
The black dotted curve represents the case where the QCR is on, but neglecting the quantum interference between the degenerate levels, that is, $\Gamma^{(1)}(\phi_0,\phi_1,\phi_1,\phi_0)=0$.
The light-blue dashed line is the theory curve with the form of $\propto \alpha^2e^{-4\alpha^2}$.
({\bf D}) Time dependence of the population of $|\phi_\alpha\rangle$ when the QCR is on (purple solid curve) and when the QCR is off (green dashed curve)\textcolor{black}{, for \textcolor{black}{$\kappa/2\pi=1.6$~kHz} and \textcolor{black}{$\gamma_p/2\pi=0.8$~kHz}}. The black thin dashed curve is for the case with $\Gamma^{(1)}(\phi_0,\phi_1,\phi_1,\phi_0)=0$.
The triangles on the top of the figure indicate the time used in panel (E).
({\bf E}) Husimi Q function, $\langle \alpha' | \rho^{\rm KPO} |\alpha'\rangle$, at different times, for the case that the QCR is off (left) and on (right). 
We used \textcolor{black}{$eV/h=40$~GHz} for panels (C--E).
}
\label{coherence_2_26_24}
\end{figure}

According to Eq.~(\ref{EOM_10_24_23}), the $\Gamma$s relevant to the change in $\rho^{\rm KPO}_{\phi_0,\phi_1}$ are
$\Gamma^{(1)}(\phi_0,\phi_1,\phi_i,\phi_j)$, $\Gamma^{(2)}(\phi_0,\phi_1,\phi_i)$, and $\Gamma^{(3)}(\phi_0,\phi_1,\phi_i)$.
In Fig.~\ref{coherence_2_26_24}(A), we present four dominant ones much greater than the others, $\Gamma^{(1)}(\phi_0,\phi_1,\phi_0,\phi_1)$, $\Gamma^{(1)}(\phi_0,\phi_1,\phi_1,\phi_0)$, $\Gamma^{(2)}(\phi_0,\phi_1,\phi_0)$, and $\Gamma^{(3)}(\phi_0,\phi_1,\phi_1)$, where the first two are positive and increase the off-diagonal element while the latter two are negative and decrease the off-diagonal element as illustrated in Fig.~\ref{coherence_2_26_24}(B). Here, $\Gamma^{(1)}(\phi_0,\phi_1,\phi_1,\phi_0)$ is the effect of the quantum interference arising from the degeneracy of $|\phi_0\rangle$ and $|\phi_1\rangle$,
and its role is to preserve the coherence of the KPO \textcolor{black}{(see subsection ``Quantum interference effect associated with the level degeneracy" in the Methods section for the definition of the interference effect)}.
Although its amplitude is smaller than the other three, its impact on the bit-flip rate is remarkable. 
Figure~\ref{coherence_2_26_24}(C) shows the bit-flip rate  (transition rate from $|\phi_\alpha\rangle$ to $|\phi_{-\alpha}\rangle$) as a function of $\alpha$.
\textcolor{black}{
We define the QCR-induced bit-flip rate as $\Gamma_{\rm b-flip}=\lim_{\Delta t\rightarrow 0}\langle\phi_{-\alpha}|\rho^{\rm KPO}(\Delta t)|\phi_{-\alpha}\rangle/\Delta t$, where $\rho^{\rm KPO}(\Delta t)$ is calculated using Eq.~(\ref{EOM_10_24_23}) with $\rho^{\rm KPO}(0)=|\phi_{\alpha}\rangle\langle\phi_{\alpha}|$. Here, $\phi_{-\alpha}\propto |\phi_0\rangle - |\phi_1\rangle$, and $\langle \phi_{\alpha} | \phi_{-\alpha}\rangle=0$.} 
If we neglect the quantum interference between the degenerate levels, i.e., let $\Gamma^{(1)}(\phi_0,\phi_1,\phi_1,\phi_0)=0$ \textcolor{black}{in Eq.~(\ref{EOM_10_24_23})}, the bit-flip rate increases with $\alpha$, and the phase-flip rate shown in Fig.~\ref{Gamma1_com2_12_11_23}(D) does not significantly dominate over the bit-flip rate, that is, the biased nature of errors is not preserved. 
On the other hand, \textcolor{black}{ if we use $\Gamma^{(1)}(\phi_0,\phi_1,\phi_1,\phi_0)$ given in Eq.~(\ref{Gamma_12_21_23})}, the bit-flip rate steeply decreases as $\alpha$ increases and scales similarly to the case of single-photon loss, that is, the bit-flip rate is proportional to $\alpha^2e^{-4\alpha^2}$ (see \textcolor{black}{subsection ``Pure dephasing and single-photon loss" in the Methods section}).
Then, the bit-flip rate is much smaller than the phase-flip rate, and therefore the biased nature of errors is preserved under operation of the QCR. 

We examine the stability of $|\phi_\alpha\rangle$ under operation of the QCR by simulating the dynamics with the initial state of $|\phi_\alpha\rangle$ \textcolor{black}{and with $eV/h=40$~GHz}.
 \textcolor{black}{We used smaller bias voltage than in Fig.~\ref{population_2_5_24} to keep the QCR-induced phase-flip rate around $10^6$ $s^{-1}$ [see Fig.~\ref{Gamma1_com2_12_11_23}(A)].}
The population of $|\phi_\alpha\rangle$, denoted by $P_{\alpha}$, is kept higher when the QCR is on than when the QCR is off as shown in Fig.~\ref{coherence_2_26_24}(D).
It is noteworthy that if we neglect the quantum interference between the degenerate levels, $P_\alpha$ decreases even more rapidly  than when the QCR is off.
\textcolor{black}{The population of qubit states is kept higher than 0.83 when the QCR is on due to the cooling effect, while it decreases approximately to 0.3 when the QCR is off as shown in Fig.~\ref{population_2_5_24}(D).}
Thus, the Kerr-cat qubit is stabilized by the energy absorption by the QCR and the quantum interference between the degenerate levels.
Figure~\ref{coherence_2_26_24}(E) shows the Husimi Q function, $\langle\alpha'|\rho^{\rm KPO}|\alpha'\rangle$, at different times. 
The Q function widely spreads when the QCR is off because of heating effect of the pure dephasing (see the result for \textcolor{black}{$t=48~\mu$s}). 
On the other hand, it is confined around $\alpha'=\pm 2$ when the QCR is on.

\textcolor{black}{
In general, the interference can occur when the system (the KPO, in our case) has degenerate energy levels that are relevant to its dynamics, which do not have to be the ground states. 
It is also noteworthy that, even if such degenerate levels exist, the interference effect can be negligible depending on the properties of the energy eigenstates. 
The properties are reflected in the QCR-transition rates via $\eta_{\phi_\mu,\phi_\nu}^{f/b,\delta m}$.
For example, in Fig.~\ref{coherence_2_26_24}(C), the difference between the QCR-induced bit-flip rates with and without the interference effect vanishes for $\alpha\ll 1$, where the pump amplitude becomes zero and the form of the qubit Hamiltonian is the same as a transmon, while the two lowest levels, $|0\rangle$ and $|1\rangle$, are still degenerate in the rotating frame.
It implies that the interference effect is negligible in this parameter regime.
}

\section*{DISCUSSION}
We have theoretically studied on-chip refrigeration for Kerr-cat qubits with a \textcolor{black}{quantum circuit refrigerator (QCR)}.
We have examined the QCR-induced deexcitations and excitations of a \textcolor{black}{Kerr parametric oscillator (KPO)} by developing a master equation.
The rate of the QCR-induced deexcitations can be controlled by more than four orders of magnitude by tuning the bias voltage across microscopic junctions.
By examining the QCR-induced bit and phase flips, we have shown that the biased nature of errors of the qubit is preserved even under operation of QCR, that is, the bit-flip rate is much smaller than the phase-flip rate.
We have found novel quantum interference in the tunneling process which occurs when the two lowest energy levels of the KPO are degenerate, and have revealed that the QCR-induced bit flip is suppressed by more than six orders of magnitude due to the quantum interference. 
Thus, QCR can serve as a tunable dissipation source which stabilizes Kerr-cat qubits mitigating unwanted heating due to the pure dephasing.
Even though we particularly consider a KPO in this paper, our theory can be applied to more general superconducting circuits.

Although studying the performance of the QCR in specific applications of KPOs is beyond the scope of this paper, we comment on two possible applications and directions for future study.
A possible application of the QCR is the stabilization of Kerr-cat qubits during gate-based quantum computing.
The QCR can reduce leakage errors into excited states which cannot be corrected by quantum error correction protocols that only deal with errors in the qubit subspace. 
The QCR may also find application in measurement-based state preparation of the Kerr-cat qubit, which was proposed in Ref.~\cite{Suzuki2022} and experimentally utilized in Ref.~\cite{Frattini2022,Venkatraman2022}. 
Homodyne and heterodyne detections can be used to determine in which side of the effective double-well potential the KPO is trapped. Therefore, the measurement can tell us that the system is in either of $|\phi_\alpha\rangle$ and $|\phi_{-\alpha}\rangle$ if the system is confined in the qubit subspace because $|\phi_\alpha\rangle$ and $|\phi_{-\alpha}\rangle$ are in opposite potential wells. 
However, the total population of the excited states gives rise to the error of the state preparation.
Because a QCR can reduce the population of excited states, activation of a QCR prior to the measurement-based state preparation would increase the fidelity of the state preparation.

\textcolor{black}{We discuss disadvantages of the use of a QCR. 
The relevant drawback of the QCR is the QCR-induced phase flip, which should be corrected for large-scale quantum computations. 
This limits the applicability of the QCR to cases where the pure dephasing rate is sufficiently smaller than the acceptable phase-flip rate, which will vary across different applications. 
The simple and wide tunability of the cooling rates of the QCR will help to adjust the cooling performance balanced against the unwanted QCR-induced phase flip.
However, the use of a QCR will degrade system coherence to an impractical level for error correction when the pure dephasing rate is too high, although a QCR may still be useful for qubit reset. 
As shown in Fig.~\ref{Gamma1_com2_12_11_23}(D), the QCR-induced phase-flip rate decreases as the size of coherent states $\alpha$ decreases. 
A possible way to mitigate the issue of the unwanted QCR-induced phase flip is to find appropriate $\alpha$ that is small enough to achieve an acceptably slow QCR-induced phase-flip rate, yet large enough to ensure practical biased noise.
}

\textcolor{black}{
As seen in Fig.~\ref{Gamma1_com2_12_11_23}(A), there is phase flip with the rate less than $10^3~{\rm s}^{-1}$ even in the absence of the bias voltage $V$, due to finite photon assisted electron tunneling. This residual phase flip can be reduced by decreasing the coupling strength between the QCR and the KPO, although this comes at the cost of reducing the maximum amplitude of the cooling rate. The heating rate at $V=0$ is less than $10~{\rm s}^{-1}$, and is therefore negligible.}

\textcolor{black}{
}

\textcolor{black}{
We summarize the pros and cons of our scheme comparing it with the previous works based on two-photon dissipation~\cite{Touzard2018,Puri2020} and frequency-selective dissipation~\cite{Putterman2022}.
(i) Both of the previous schemes utilize additional resonators. In contrast, our scheme uses a SINIS junction which is significantly smaller in size compared to the resonators.
(ii) The frequency-selective dissipation does not require any additional drives. The two-photon dissipation is activated by a microwave applied to the additional resonator, while the QCR is driven by a DC bias voltage across the junction.
(iii) The QCR is insensitive to parameters of the KPO, such as resonance frequency, nonlinearity parameter, and pump amplitude, which is a useful feature for scaling up the system.
In contrast, the frequency of the microwave used for the two-photon dissipation depends on the resonance frequencies of both the qubit and the additional resonator. 
For the frequency-selective dissipation, the resonance frequencies of the additional resonators  must be nearly identical, and these frequencies are determined by the parameters of the KPO.
(iv) The QCR also functions for relatively small values of $\alpha$, e.g., $\alpha=2$, where the frequency-selective dissipation tends to increase bit flip errors compared to the case without the dissipation mechanism~\cite{Putterman2022}.
(v) The advantage of the previous schemes is that phase flip is not enhanced in an ideal situation, whereas our scheme induces phase flip. 
}

\textcolor{black}{
In this paper, we neglected the Johnson-Nyquist noise from the normal metal island of the QCR, while the electron temperature of the normal metal island was accounted for in the calculation of the QCR-induced transition rates via the Fermi-Dirac distribution function. 
Although the Johnson-Nyquist noise could potentially affect properties of the attached resonator and qubit, such an effect has not yet been observed in previous QCR measurements~\cite{Tan2017,Silveri2019,Yoshioka2021,Sevriuk2022,Yoshioka2023}. 
We attribute this to the fact that the electron temperature of the normal metal island at the voltage used for cooling is lower than that at $V=0$, due to the tunneling of high-energy electrons enhanced by the bias voltage~\cite{Lowell2013,Tan2017}, and that the volume of the normal meal island is small, typically on the order of 0.01~$\mu{m}^3$~\cite{Tan2017}.
}

\textcolor{black}{
In our derivation of the QCR-induced transition rates, the normal metal island is set in a stationary state determined by the elastic electron tunneling. This is based on assumptions that the dynamics of the normal metal island is governed by elastic electron tunneling which is much faster than the photon assisted electron tunneling for the parameters used, and thus the stationary state determined by the elastic electron tunneling provides a good approximation for the state of the normal metal island.
In Ref.~\cite{Hsu2020}, the authors studied the charge dynamics of the normal metal island under the operation of the QCR, and showed that the effect of the charge dynamics on the qubit reset is limited for typical parameters.
They also clarified that when the size of the normal metal island is much smaller and enters in the quantum dot regime, the dynamics of the normal metal island becomes significant, leading to the emergence of different phenomena. Studying charge dynamics with our system will be an interesting direction for future research.
}

\section*{METHODS}
\label{Methods}

\subsection*{Unitary transformations}
\label{SM: Unitary transformation}
We apply unitary transformations $U_j$, $U$ and $U_{\rm RF}$ to simplify the Hamiltonian and to move into a rotating frame at frequency of $\omega_p/2$. 
We begin with $U_j$ defined by
\begin{eqnarray}
U_j  = \exp\Big{[}\frac{i}{\hbar} Q_j \Phi_N\Big{]},
\end{eqnarray} 
which satisfies $U_j  (Q_N + Q_j) U_j^\dagger = Q_N$.
The unitary transformation $U_j$ simplifies $H_0$ by eliminating $Q_j$ as
\begin{eqnarray} 
H_{0} &=& \frac{Q_N^2}{2C_N} + \frac{(Q+ \alpha_{c} Q_N)^2}{2C_r} - E_J \cos\Big{(}\frac{2e}{\hbar}\Phi\Big{)}.
\end{eqnarray} 
Because there is no $\Phi_N$ in the Hamiltonian we can further rewrite it as 
\begin{eqnarray} 
H_{0} &=& \sum_q \Big{[}\frac{e^2q^2}{2C_N} + \frac{(Q+ \alpha_{c} e q)^2}{2C_r} - E_J \cos\Big{(}\frac{2e}{\hbar}\Phi\Big{)}\Big{]} |q\rangle \langle q|,
\label{H0_12_19_23}
\end{eqnarray} 
where $q$ is an integer denoting the number of the excess charge in the normal-metal island, that is, $eq$ is the charge in the normal-metal island.
Note that $U_j$ does not change $H_{\rm QP}$ and  $H_{T}$.

Next, we perform a unitary transformation 
\begin{eqnarray} 
U = \sum_q \exp \Big{[} \frac{i}{\hbar}  \alpha_{c} e q \Phi \Big{]} |q\rangle \langle q|
\label{Uq_8_2_23}
\end{eqnarray}
to simplify $H_0$ in Eq.~(\ref{H0_12_19_23}) to  
\begin{eqnarray} 
H_{0} &=& \sum_q \Big{[} \frac{e^2q^2}{2C_N} + \frac{Q^2}{2C_r} - E_J \cos\Big{(}\frac{2e}{\hbar}\Phi\Big{)}\Big{]} |q\rangle \langle q|,
\end{eqnarray}
by eliminating $ \alpha_{c} e q$ from the second term, where we used the fact that $U_q= \exp [\frac{i}{\hbar}  \alpha_{c} e q \Phi]$ translates the charge operator as
$U_q (Q+ \alpha_{c} e q) U_q^\dagger = Q$.
The operator $U$ changes $H_T$ while $H_{\rm QP}$ is unchanged.
The effect of $U$ on $H_T$ is discussed later.
We  further rewrite $H_0$ by using $\phi = \frac{2e}{\hbar}\Phi$ and $n=Q/2e$  as
\begin{eqnarray} 
H_{0}  &=&  \sum_q \Big{[} \frac{e^2q^2}{2C_N} + 4E_C n^2  - E_J(t) \cos\phi \Big{]} |q\rangle \langle q|,\nonumber\\
&=& \sum_q \Big{[} \frac{e^2q^2}{2C_N}  |q\rangle \langle q|  \Big{]} + H_{\rm KPO}(t),
\label{H0_8_7_23}
\end{eqnarray} 
where $[\phi,n]=i$, and $H_{\rm KPO}(t)$ is the Hamiltonian of the KPO defined by
\begin{eqnarray} 
H_{\rm KPO}(t) &=& 4E_C n^2 - E_J(t) \cos\phi.
\end{eqnarray} 

We focus on the Hamiltonian of the KPO, $H_{\rm KPO}$.
The magnetic flux \textcolor{black}{$\Phi(t)$} in the SQUID is harmonically modulated around its mean value with small amplitude\textcolor{black}{.
$E_J(t)$ is represented as $E_J(t)=\bar{E}_J \cos(\pi\Phi(t)/\Phi_0)$ where $\bar{E}_J$ is constant. We assume that  $\Phi(t)=\Phi_{\rm dc}-\delta_p\Phi_0\cos(\omega_p t)$, where $\Phi_{\rm dc}$, and $\delta_p(\ll 1)$ are constant. Then, $E_J(t)$ can be approximated as $E_J+\delta E_J \cos(\omega_p t)$, where $E_J=\bar{E}_J\cos(\pi\Phi_{\rm dc}/\Phi_0)$ and $\delta E_J = \bar{E}_J\pi\delta_p\sin(\pi\Phi_{\rm dc}/\Phi_0)$ (see, e.g., section 4.1 of \cite{Goto2019}).}
The Taylor expansion leads to 
\begin{eqnarray}
H_{\rm KPO}(t) &=& 4E_C n^2 - E_J \Big{(}
1 - \frac{1}{2} \phi^2 
+ \frac{1}{24}\phi^4 + \cdots \Big{)} \nonumber\\
&& - \delta E_J \Big{(} 1  - \frac{1}{2} \phi^2
+ \frac{1}{24}\phi^4 + \cdots \Big{)} \cos(\omega_p t).
\label{H_KPO_8_7_23}
\end{eqnarray}
The quadratic time-independent part of the \textcolor{black}{Hamiltonian} (\ref{H_KPO_8_7_23})
can be diagonalized by using relations
$n = -in_0(a-a^\dagger)$ and $\phi = \phi_0 (a+a^\dagger)$,
where $n_0^2=\sqrt{E_J/(32 E_C)}$ and 
$\phi_0^2 = \sqrt{2E_C/E_J}$ are the zero-point fluctuations.
Taking into account up to the 4th order of $\phi$, we obtain 
\textcolor{black}{
\begin{eqnarray}
\frac{H_{\rm KPO}(t)}{\hbar} &=& \omega_c^{(0)} \Big{(} a^\dagger a + \frac{1}{2} \Big{)}
- \frac{\chi}{12} (a + a^\dagger)^4\nonumber\\
&& + \Big{[} - \frac{\delta E_J}{\hbar} +
 2\beta (a + a^\dagger)^2 
- \frac{2\chi \beta}{3\omega_c^{(0)}} (a + a^\dagger)^4 \Big{]}
\cos(\omega_p t),
\end{eqnarray}}
where we have defined the resonance frequency $\omega_c^{(0)} =\frac{1}{\hbar} \sqrt{8E_CE_J}$, the Kerr nonlinearity $\chi=E_C/\hbar $,
the parametric drive strength $\beta = \omega_c^{(0)} \delta E_J/ (8E_J)$.
We neglect the last term in the square brakets because it is much smaller than the other terms ($\chi\beta\ll  \omega_c^{(0)}$).
We also drop c-valued terms in the expression above and obtain
\begin{eqnarray}
H_{\rm KPO}(t) / \hbar &=& \omega_c^{(0)} a^\dagger a 
- \frac{\chi}{12} (a + a^\dagger)^4 + 2\beta (a + a^\dagger)^2
\cos(\omega_p t).
\end{eqnarray}

Now, we move into a rotating frame at the frequency
$\omega_p/2$ by transforming the system with unitary operator 
\begin{eqnarray}
U_{\rm RF}(t) = e^{i\frac{\omega_{p}}{2} t a^\dagger a} \otimes I_N = \sum_m  e^{i\frac{m\omega_p t}{2} } |m\rangle \langle m| \otimes I_N,
\label{URF_5_1_23}
\end{eqnarray}
where $I_N = \sum_q |q\rangle \langle q|$.
After the unitary transformation the KPO Hamiltonian is written as
\begin{eqnarray}
\frac{H_{\rm KPO}^{\rm (RF)}(t)}{  \hbar} &=& (\omega_c^{(0)}-\omega_p/2) a^\dagger  a 
- \frac{\chi}{12} (a e^{-i\frac{\omega_p}{2}t} + a^\dagger e^{i\frac{\omega_p}{2}t} )^4 + 2\beta (a e^{-i\frac{\omega_p}{2}t}  + a^\dagger e^{i\frac{\omega_p}{2}t})^2
\cos(\omega_p t).
\label{H_KPO_RF_11_25_2}
\nonumber\\
\end{eqnarray}
To obtain \textcolor{black}{Eq.~(\ref{HKPO_11_15_24})}, we use the rotating-wave approximation,
which is valid when $|\omega_c^{(0)} - \omega_p/2|$, 
$\chi/12$, and $2\beta$ are all much 
smaller than $2\omega_p$.
The detuning $\Delta_{\rm KPO}$ in \textcolor{black}{Eq.~(\ref{HKPO_11_15_24})} is given by $\Delta_{\rm KPO} = \omega_c - \omega_p/2$, where $ \omega_c$ is the dressed resonator frequency defined by $ \omega_c = \omega_c^{(0)}-\chi$.
\textcolor{black}{The term proportional to $\beta a^\dagger a \cos(\omega_p t)$ in Eq.~(\ref{H_KPO_RF_11_25_2}) is omitted in the rotating wave approximation, and therefore the dressed resonator frequency is independent of $\beta$.}


We consider the effect of $U$ and $U_{\rm RF}$ on $H_T$.
The unitary operators transform $H_T$ as 
\begin{eqnarray}
H_T^{\rm (RF)} &=& U_{\rm RF} U H_T U^\dagger U_{\rm RF}^\dagger \nonumber \\
&=& \sum_{m,m'} \sum_{k,l,\sigma} \sum_{q,q'} e^{i\omega_{\rm RF} t (m'-m) t} 
\langle m'|  \exp\Big{[} \frac{i}{\hbar}  \alpha_{c} e (q'-q) \Phi \Big{]}  |m\rangle \nonumber \\
&& \times  ( \langle q'|  e^{-i\frac{e}{\hbar} \Phi_N}  |q\rangle T_{lk} d_{l\sigma}^\dagger c_{k\sigma} 
+ \langle q'|  e^{i\frac{e}{\hbar} \Phi_N}  |q\rangle T_{lk}^\ast c_{k\sigma}^\dagger  d_{l\sigma} )
 \times  |q',m'\rangle \langle q,m|\nonumber\\
&=& \sum_{m,m'} \sum_{k,l,\sigma} \sum_{q} e^{i\omega_{\rm RF} t (m'-m) t} \Big{[}
\langle m'|  \exp\Big{[} -\frac{i}{\hbar}  \alpha_{c} e \Phi \Big{]}  |m\rangle T_{lk} d_{l\sigma}^\dagger c_{k\sigma} |q-1,m'\rangle \langle q,m| \nonumber\\
&& + \langle m'|  \exp\Big{[} \frac{i}{\hbar}  \alpha_{c} e \Phi \Big{]}  |m\rangle
T_{lk}^\ast c_{k\sigma}^\dagger  d_{l\sigma} |q+1,m'\rangle \langle q,m| \Big{]}.
\label{H_T_8_7_23_2}
\end{eqnarray}
In the above equation we used the following fact.
Because the unitary operator $U_e=e^{i\frac{e}{\hbar} \Phi_N}$
shifts the charge state as 
\begin{eqnarray}
U_e |q\rangle = |q+1\rangle,
\label{U_e_5_1_23}
\end{eqnarray}
we have
\begin{eqnarray}
\langle q| e^{-i\frac{e}{\hbar} \Phi_N} |q'\rangle &=& \langle q| U_e^\dagger |q'\rangle \nonumber\\
&=&  \langle q+1 |q'\rangle.
\label{q_5_1_23}
\end{eqnarray}
In the derivation of Eq.~(\ref{U_e_5_1_23}), we used 
\begin{eqnarray}
U_e Q_N U_e^\dagger = Q_N - e.
\end{eqnarray}
The second term in Eq.~(\ref{H_T_8_7_23_2}) corresponds to the electron tunneling from the normal-metal island to the superconducting electrode (note that the positive charge in the normal-metal island increases because of this transition).
By using $\delta m = m'-m$ in Eq.~(\ref{H_T_8_7_23_2}), we can obtain Eq.~(\ref{HRF_2_12_24}).

\textcolor{black}{\subsection*{Pure dephasing and single-photon loss}}
We consider a \textcolor{black}{Kerr parametric oscillator (KPO)} without a \textcolor{black}{superconductor--insulator--normal-metal--insulator--superconductor (SINIS)} junction.
The master equation of the KPO is given by
\begin{eqnarray}
\frac{d\rho(t)}{dt} = - \frac{i}{\hbar} [H_{\rm KPO}^{\rm (RF)},\rho(t)]  
+ \frac{\kappa}{2} \mathcal{D}[a] \rho(t)
+ \gamma_p \mathcal{D}[a^\dagger a] \rho(t),
\label{ME_wo_QCR_4_8_24}
\end{eqnarray} 
where $\mathcal{D}[\hat{O}]\rho = 2\hat{O}\rho \hat{O}^\dagger -  \hat{O}^\dagger \hat{O} \rho - \rho \hat{O}^\dagger \hat{O}$~\cite{Puri2020}. Here, $\kappa$ and $\gamma_p$ are the single-photon-loss rate and the pure-dephasing rate, respectively.
We define the transition rate from $|\psi_i\rangle$ to $|\psi_f\rangle$ due to the pure dephasing as 
\begin{eqnarray}
\Gamma_{p}^{i\rightarrow f} = \gamma_p \langle \psi_f | \mathcal{D}[a^\dagger a] \rho_i |\psi_f\rangle,
\end{eqnarray} 
where $ \rho_i = |\psi_i\rangle\langle \psi_i|$ and $\langle \psi_f|\psi_i\rangle=0$.

\begin{figure}
\begin{center}
\includegraphics[width=14cm]{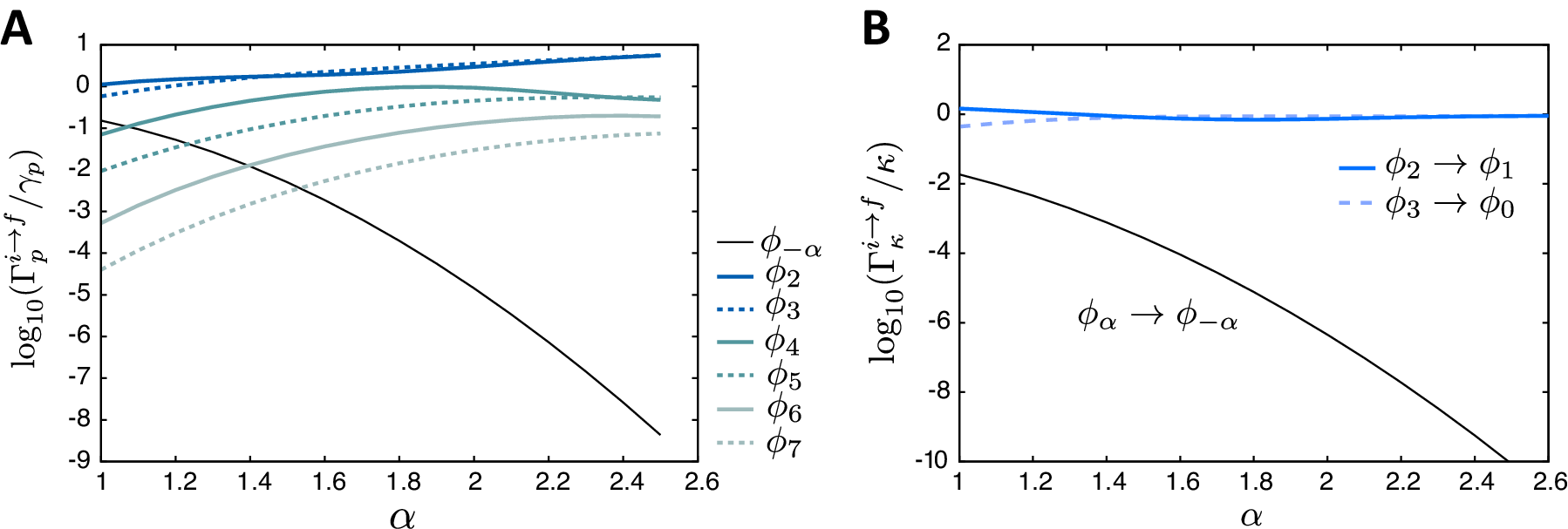}
\end{center}
\caption{
({\bf A}) Rate of transitions from $|\phi_\alpha\rangle$ to other states due to the pure dephasing for $\Delta_{\rm KPO}=0$.
The rate is normalized by $\gamma_p$.
({\bf B}) Rate of deexcitations and bit flip caused by single-photon loss, where the rate is normalized by $\kappa$.
}
\label{rates_gamma_kappa_4_16_24}
\end{figure}

We examine the transition rates from $|\psi_i\rangle=|\phi_\alpha\rangle$ to other states.
The transition rates normalized by $\gamma_p$ are presented for different final states in Fig.~\ref{rates_gamma_kappa_4_16_24}(A).
The bit-flip rate (transition from $|\phi_\alpha\rangle$ to $|\phi_{-\alpha}\rangle$) is suppressed as $\alpha$ increases, and is explicitly written as
\begin{eqnarray}
\Gamma_{p}^{\phi_\alpha \rightarrow \phi_{-\alpha}}/\gamma_p &=& 2\alpha^4 [(x^2+y^2) e^{-2\alpha^2} - 2xy]^2 \nonumber \\
&& 
- 2\Big{[} 
\alpha^2 \Big\{ -(x^2+y^2) e^{-2\alpha^2} + 2xy \Big\} + \alpha^4 \Big\{ (x^2+y^2) e^{-2\alpha^2}
+ 2xy \Big\} \Big{]} \nonumber \\
&& \times \Big{[}(x^2+y^2) e^{-2\alpha^2} + 2xy\Big{]},
\label{Gammap_5_24_24}
\end{eqnarray} 
with $x=\frac{1}{\sqrt{2}}(N_++N_-)$ and $y=\frac{1}{\sqrt{2}}(N_+-N_-)$.
$\Gamma_{p}^{\phi_\alpha \rightarrow \phi_{-\alpha}}/\gamma_p$ in Eq.~(\ref{Gammap_5_24_24}) is approximated by $2\alpha^2e^{-4\alpha^2}$ for sufficiently large $|\alpha|$.
The transition rates outside the qubit subspace become much larger than the bit-flip rate as $\alpha$ increases.
Especially transition rates to the adjacent excited states $|\phi_{2,3}\rangle$ are higher than $\gamma_p$ itself for $\alpha>1.4$.

Similarly, we define the transition rate from $|\psi_i\rangle$ to $|\psi_f\rangle$ due to the single-photon loss as 
\begin{eqnarray}
\Gamma_{\kappa}^{i\rightarrow f} = \kappa \langle \psi_f | \mathcal{D}[a] \rho_i |\psi_f\rangle.
\end{eqnarray} 
Figure~\ref{rates_gamma_kappa_4_16_24}(B) shows the rate of relevant deexcitations to qubit states and bit flip caused by the single-photon loss. 
The bit-flip rate is suppressed as $\alpha$ increases, and is explicitly written as 
\begin{eqnarray}
\Gamma_\kappa^{\phi_\alpha\rightarrow \phi_{-\alpha}}/\kappa &=& (x^2-y^2)^2 \alpha^2 e^{-4\alpha^2} \nonumber\\
&& + \alpha^2\Big{(} (x^2+y^2) e^{-2\alpha^2} - 2xy \Big{)}
\Big{(}(x^2+y^2) e^{-2\alpha^2} + 2xy\Big{)},
\end{eqnarray} 
which is well approximated by $2\alpha^2 e^{-4\alpha^2}$ for sufficiently large $|\alpha|$.
The deexcitation rates asymptotically approach $\kappa$, that is, $\Gamma_\kappa^{\phi_{0(1)}\rightarrow \phi_{3(2)}} \rightarrow \kappa$, which is derived by using $|\phi_{2,3}\rangle\simeq
\frac{1}{\sqrt{2}}({D}(\alpha)|1\rangle \mp {D}(-\alpha)|1\rangle)$ for sufficiently large $|\alpha|$,
where $D(\alpha)$ is the displacement operator defined by $D(\alpha)=\exp[\alpha a^\dagger - \alpha^\ast a]$.

\textcolor{black}{\subsection*{Derivation of master equation}}
Suppose that at time $t$ the state of the total system is given by 
\begin{eqnarray} 
|\Psi(t)\rangle = \sum_\mu a_\mu(t) |\psi_\mu\rangle.
\end{eqnarray}  
The time evolution of the total system is governed by the Schr\"{o}dinger equation, 
\begin{eqnarray} 
\frac{\partial}{\partial t}a_\mu(t) = -\frac{i}{\hbar} E_\mu a_\mu(t) - \frac{i}{\hbar} \sum_\nu V_{\mu\nu}(t) a_\nu(t),
\label{EOM1_12_20_23}
\end{eqnarray} 
where $V_{\mu\nu}(t) = \langle \psi_\mu| H_T^{\rm (RF)}(t)|\psi_\nu\rangle$.
\textcolor{black}{Integrating Eq.~(\ref{EOM1_12_20_23}) over time leads to the integral equation, 
\begin{eqnarray} 
a_\mu(t) = e^{-iE_\mu t/\hbar}a_\mu(0) - \frac{i}{\hbar} \sum_\nu \int_0^t ds e^{-iE_\mu (t-s)/\hbar} V_{\mu\nu}(s) a_\nu(s).
\label{am_12_20_23}
\end{eqnarray} 
The validity of this equation can be easily confirmed by differentiating the equation with respect to time.
Because of Eq.~(\ref{am_12_20_23}) we have
\begin{eqnarray} 
a_\nu(s) = e^{-iE_\nu s/\hbar}a_\nu(0) - \frac{i}{\hbar} \sum_{\xi} \int_0^s ds' e^{-iE_\nu (s-s')/\hbar} V_{\nu\xi}(s') a_{\xi}(s').
\label{an_12_20_23}
\end{eqnarray} 
We use Eq.~(\ref{an_12_20_23}) in the right-hand side of Eq.~(\ref{am_12_20_23}) and repeat the same procedure to obtain} the solution to second order in the perturbation as
\begin{eqnarray} 
a_\mu(t) &\simeq& e^{-iE_\mu t/\hbar}a_\mu(0) - \frac{i}{\hbar} \sum_\nu \int_0^t ds e^{-iE_\mu (t-s)/\hbar} V_{\mu\nu}(s) e^{-iE_\nu s/\hbar}a_\nu(0) \nonumber\\
&& - \frac{1}{\hbar^2} \sum_{\nu,\xi} 
\int_0^t ds  \int_0^s ds' V_{\mu\nu}(s)  V_{\nu\xi}(s')
e^{-iE_\mu t/\hbar} e^{i\omega_{\mu\nu}s} e^{i\omega_{\nu\xi}s'} a_\xi(0).
\label{am_2_12_24}
\end{eqnarray} 
As seen from Eq.~(\ref{HRF_2_12_24}), the perturbation can be written as $V_{\mu\nu}(t) = \sum_{\delta m}V_{\mu\nu}^{(\delta m)} \exp[i\omega_{\rm RF}\delta m t]$ with $V_{\mu\nu}^{(\delta m)} = \langle \psi_\mu | V^{(\delta m)} | \psi_\nu\rangle$, where
\begin{eqnarray}
V^{(\delta m)}  &=& \sum_{m} \sum_{k,l,\sigma} \sum_{q}   
\Big\{
\langle \delta m+m|  \exp\Big{[} - \frac{i}{\hbar}  \alpha_{c} e \Phi \Big{]}   |m\rangle
\cdot T_{lk} d_{l\sigma}^\dagger c_{k\sigma}  |q-1,\delta m+m\rangle \langle q,m|   
 \nonumber\\
&& + \langle \delta m+m|  \exp\Big{[} \frac{i}{\hbar}  \alpha_{c} e \Phi \Big{]}  |m\rangle
\cdot T_{lk}^\ast c_{k\sigma}^\dagger  d_{l\sigma}
 |q+1,\delta m+m\rangle \langle q,m|  \Big\}. 
\label{V_12_22_23}
\end{eqnarray}
By using Eq.~(\ref{am_2_12_24}), we can write elements of the density matrix, $a_\mu(t) a_{\mu'}^\ast(t)$, as  
\begin{eqnarray} 
 \frac{a_\mu(t) a_{\mu'}^\ast (t)}{e^{-i\omega_{\mu\mu'}t}} 
&\simeq&  a_\mu(0) a_{\mu'}^\ast(0) + \frac{i}{\hbar}  \sum_{\nu'} 
\sum_{\delta m'} (V_{\mu' \nu'}^{(\delta m')})^\ast  \int_0^t  ds 
e^{-i(\omega_{\mu'\nu'} + \omega_{\rm RF} \delta m')s} 
a_\mu(0) a_{\nu'}^\ast(0)
 \nonumber\\
&&- \frac{i}{\hbar}  \sum_{\nu} 
\sum_{\delta m} V_{\mu \nu}^{(\delta m)}  \int_0^t  ds 
e^{i(\omega_{\mu\nu} + \omega_{\rm RF} \delta m)s} 
a_\nu(0) a_{\mu'}^\ast(0)
 \nonumber\\
&& + \frac{1}{\hbar^2}  \sum_{\nu,\nu'} \sum_{\delta m, \delta m'}  V_{\mu \nu}^{(\delta m)} (V_{\mu' \nu'}^{(\delta m')})^\ast 
\int_0^t  ds e^{i(\omega_{\mu\nu} + \omega_{\rm RF} \delta m)s} 
 \int_0^t  ds e^{-i(\omega_{\mu'\nu'} + \omega_{\rm RF} \delta m')s}  a_\nu(0) a_{\nu'}^\ast(0)\nonumber\\
&& - \frac{1}{\hbar^2} \sum_{\nu,\xi}  \sum_{\delta m, \delta m'} (V_{\mu'\nu}^{(\delta m)})^\ast V_{\xi\nu}^{(\delta m')} \int_0^t  ds  \int_0^s ds'
e^{-i(\omega_{\mu'\nu}+\omega_{\rm RF}\delta m)s} e^{i(\omega_{\xi\nu}+\omega_{\rm RF}\delta m')s'} a_\mu(0) a_\xi^\ast(0) \nonumber\\
&& - \frac{1}{\hbar^2} \sum_{\nu,\xi}  \sum_{\delta m, \delta m'} V_{\mu\nu}^{(\delta m)}(V_{\xi\nu}^{(\delta m')})^\ast  \int_0^t  ds  \int_0^s ds'
e^{i(\omega_{\mu\nu} + \omega_{\rm RF}\delta m) s} e^{-i(\omega_{\xi\nu} + \omega_{\rm RF}\delta m')s'} a_{\xi}(0) a_{\mu'}^\ast(0).
\nonumber\\
\label{am_12_18_23}
\end{eqnarray}
The first term represents the evolution of the density matrix without the perturbation, the other terms represent the perturbation effects and include contributions from other elements of the density matrix.
By using Eq.~(\ref{am_12_18_23}) and the results 
\textcolor{black}{of subsection ``Time integrals in Eq.~(\ref{am_12_18_23})" in the Methods section}, we obtain 
\begin{eqnarray}
&& \frac{a_\mu(t+\Delta t) a_{\mu'}^\ast (t+\Delta t)}{e^{-i\omega_{\mu\mu'}\Delta t}} =  a_{\mu}(t) a_{\mu'}^\ast(t)\nonumber\\
&& \hspace{1cm}
+  \frac{\pi \Delta t}{\hbar} \sum_\nu 
 \sum_{\delta m,\delta m'} 
\Big{[} 
2 \sum_{\phi_{\nu'}}'  V_{\mu\nu}^{(\delta m)} (V_{\mu'\nu'}^{(\delta m')})^\ast
\delta(E_{\phi_{\mu}}  + E_{\mathcal{E}_{\mu}} - E_{\phi_\nu} - E_{\mathcal{E}_{\nu}} + \hbar\omega_{\rm RF}\delta m) a_\nu(t) a_{\nu'}^\ast(t) 
\nonumber\\
&& \hspace{1cm} -   \sum_{\phi_\xi}'  (V_{\nu\mu}^{(\delta m)})^\ast V_{\nu\xi}^{(\delta m')}
\delta(E_{\phi_\nu}  + E_{\mathcal{E}_{\nu}} - E_{\phi_{\mu}} - E_{\mathcal{E}_{\mu}} + \hbar\omega_{\rm RF}\delta m) a_\xi(t) a_{\mu'}^\ast(t)
\nonumber\\
&&\hspace{1cm} - \sum_{\phi_\xi}''  V_{\nu\mu'}^{(\delta m)} (V_{\nu\xi}^{(\delta m')})^\ast 
\delta(E_{\phi_{\nu}}  + E_{\mathcal{E}_{\nu}} - E_{\phi_{\mu'}} - E_{\mathcal{E}_{\mu}} + \hbar\omega_{\rm RF}\delta m)
 a_\mu(t) a_{\xi}^\ast(t)   \Big{]},
 \label{aa_2_14_24}
\end{eqnarray} 
where $\sum_{\phi_{\nu'}}'$, $\sum_{\phi_\xi}'$, and $\sum_{\phi_\xi}''$, respectively, denote the summation with respect to $\phi_{\nu'}$, $\phi_{\xi}$, and $\phi_{\xi}$
which satisfy Eq.~(\ref{matching_12_21_23}).
Note that in Eq.~(\ref{aa_2_14_24}), we have replaced the time interval $t$ by $\Delta t$ and shifted the origin of time by $t$.

The KPO master equation in Eq. (\ref{EOM_10_24_23}) can be obtained by tracing out the environment from Eq.~(\ref{aa_2_14_24}). 
The derivation is based on the following assumptions: At time $t$, the density matrix is represented as $\rho(t)=\rho_{\rm sys}(t) \otimes \rho_{\rm env}^{(0)}$. Here, $\rho_{\rm env}^{(0)}$ is a thermal state of the environment written as $\rho_{\rm env}^{(0)}=\sum_\mathcal{E} p_\mathcal{E} |\mathcal{E}\rangle \langle\mathcal{E}|$ with energy eigenstates $|\mathcal{E}\rangle$, where $p_\mathcal{E}$ is the probability that the state of the environment is $|\mathcal{E}\rangle$. 
At time $t+\Delta t$, $\rho(t+\Delta t)$ cannot be written as a product state of the system and the environment in general. We assume that the environment relaxes to the original state $\rho_{\rm env}^{(0)}$ in time much shorter than $\Delta t$ and the system can be represented again in a product state as $\rho(t+\Delta t)= \rho_{\rm sys}(t+\Delta t)\otimes\rho_{\rm env}^{(0)} $ where $\rho_{\rm sys}(t+\Delta t)={\rm Tr_{env}} \rho(t+\Delta t)$.
This process repeats at each time step $\Delta t$. 

To obtain $\rho^{\rm KPO}_{\phi_{\mu},\phi_{\mu'}}(t+\Delta t)$ we calculate 
$\sum_\mathcal{E_\mu} a_\mu(t+\Delta t) a_{\mu'}^\ast (t+\Delta t)$ using Eq.~(\ref{aa_2_14_24}), where $|\mathcal{E_{\mu'}}\rangle=|\mathcal{E_\mu}\rangle$.
As an example, we consider the contribution of the term with $a_\nu(t)a_{\nu'}^\ast(t)$ in Eq.~(\ref{aa_2_14_24})
particularly focusing on the term including $c_{k\sigma}^\dagger d_{l\sigma}|q+1,\delta m+m\rangle \langle q,m|$ of $V^{(\delta m)}$ in Eq.~(\ref{V_12_22_23}).
For deriving the contribution of the term to $\rho^{\rm KPO}_{\phi_{\mu},\phi_{\mu'}}(t+\Delta t)$, we note the following points: (a) we consider only the case where $|\mathcal{E}_\nu\rangle=|\mathcal{E}_{\nu'}\rangle$ because $a_\nu(t)a_{\nu'}^\ast(t)=0$ otherwise; (b) the summation $\sum_{k,l,\sigma}$ is represented as $2\int d\varepsilon_k\int d\varepsilon_l n_s(\varepsilon_k)$, where $n_s$ is the density of state of the \textcolor{black}{quasiparticle}s in the superconducting electrode; (c) summation $\sum_\nu$ can be represented as $\sum_{\mathcal{E}_\nu}\sum_{\phi_\nu}$, and summations $\sum_\mathcal{E_\mu}\sum_{\mathcal{E}_\nu}$ are unified to $\sum_{{\rm QP}_{kl}}$ which represents the sum running over the state of \textcolor{black}{quasiparticle}s except for modes $k$ and $l$ because the state of modes $k$ and $l$ and the normal metal are determined by $c_{k\sigma}^\dagger d_{l\sigma}|q+1,\delta m+m\rangle \langle q,m|$, while the states of other \textcolor{black}{quasiparticle} modes should be the same between $|\mathcal{E}_\mu\rangle$ and $|\mathcal{E}_\nu\rangle$; (d) $\sum_{{\rm QP}_{kl}} a_\nu(t) a_{\nu'}(t)$ leads to the factor $p_q[1-f(\varepsilon_k,T_S)]f(\varepsilon_l,T_N)\rho_{\phi_\nu, \phi_{\nu'}}^{\rm KPO}(t)$.
Taking into account these points, we find that the contribution from $\rho_{\phi_\nu, \phi_{\nu'}}^{\rm KPO}(t)$ to $\rho^{\rm KPO}_{\phi_{\mu},\phi_{\mu'}}(t+\Delta t)$ is written as 
\begin{eqnarray}
\frac{{4}\pi|T|^2 \Delta t}{\hbar} \sum_{\delta m, \delta m', q} \sum_{\phi_\nu}\sum_{\phi_{\nu'}}' \int d\varepsilon_k n_s(\varepsilon_k)
p_q[1-f(\varepsilon_k,T_S)]f(\varepsilon_l^{(f,\delta m,1)},T_N)
\eta_{\phi_{\mu},\phi_{\nu}}^{(\delta m, f)} (\eta_{\phi_{\mu'},\phi_{\nu'}}^{(\delta m', f)})^\ast
\rho_{\phi_{\nu}, \phi_{\nu'}}^{\rm KPO}(t).\nonumber\\
\end{eqnarray} 
The contributions from the other terms and another NIS junction to $\rho^{\rm KPO}_{\phi_{\mu},\phi_{\mu'}}(t+\Delta t)$ can be calculated in the same manner, and thus Eq.~(\ref{EOM_10_24_23}) is obtained.
In Eq.~(\ref{Gamma_12_21_23}), we replaced $4\pi|T|^2/\hbar$ by $1/e^2 R_T$ so that 
the tunnel resistance matches the measured one for sufficiently large $V$~\cite{Ingold1992}.
The effects of the two NIS junctions are the same because they are identical in our model.

\textcolor{black}{
A comment on the derivation of the reduced master equation is in order. 
Although we considered a pure state in Eq.~(\ref{EOM1_12_20_23}) when deriving the reduced master equation, the  state at $t$ should be regarded as a mixed state of such pure states.
In our theory, the probability that the state of the system is each pure state is accounted for by factors such as $p_q$ and the Fermi-Dirac distribution function.
} 

\textcolor{black}{\subsection*{Time integrals in Eq.~(\ref{am_12_18_23})}}
We consider the time integrals in Eq.~(\ref{am_12_18_23}).
First, we consider the integral
\begin{eqnarray} 
Y_1(\omega,t) = \int_0^t e^{i(\omega-\omega_1) s}ds\int_0^t e^{-i(\omega-\omega_2) s}ds,
\label{delta1p_2_13_24}
\end{eqnarray} 
included in the fourth term of Eq.~(\ref{am_12_18_23}),
where $\omega_1$ and $\omega_2$ are constant.
$|Y_1(\omega,t)|$ peaks at $\omega=\omega_1$ and $\omega_2$.
In this study we consider the cases in which either $\omega_1=\omega_2$ or two peaks are well separated.

When $\omega_1=\omega_2$, the height of the peak is $t^2$, while its width is of the order of $2\pi/t$~\cite{Zwanzig2001}.
It is known that, for sufficiently large $t$, $Y_1$ approaches a delta function, that is,
\begin{eqnarray} 
Y_1(\omega,t)\rightarrow 2\pi t \delta(\omega-\omega_1) =  2\pi \hbar t \delta (E-E_1),
\end{eqnarray} 
where $E_1 = \hbar \omega_1=\hbar \omega_2$.
On the other hand, the integral can be neglected when two peaks are well separated because the height of the peaks are of the order of $\sqrt{\delta(E)}$.

Next, we consider 
\begin{eqnarray} 
Y_2(\omega,t) = \int_0^t  \int_0^s ds ds' e^{i(\omega-\omega_1) s} e^{-i(\omega-\omega_2) s'},
\label{delta2_2_14_24}
\end{eqnarray} 
included in the fifth and sixth terms in Eq.~(\ref{am_12_18_23}).
$|Y_2(\omega,t)|$ peaks at $\omega=\omega_1$ and $\omega_2$.
The integral can be neglected when two peaks are well separated because the height of the peaks are of the order of $\sqrt{\delta(E)}$.
When $\omega_1=\omega_2$, $Y_2(\omega,t)$ approaches a function represented by
\begin{eqnarray} 
Y_2(\omega,t)\rightarrow \pi \hbar t \delta (E-E_1) + ig(E-E_1),
\end{eqnarray} 
where the imaginary part, $ig$, can be neglected because it is an odd function about $E=E_1$ and the width becomes very narrow as $t$ increases.
By using these relations we can obtain Eq.~(\ref{aa_2_14_24}) from Eq.~(\ref{am_12_18_23}).
The condition for two peaks to be considered sufficiently separated is that
$|\omega_1-\omega_2|$ is sufficiently larger than $2\pi/t$,
because the width of each peak is of the order of $2\pi/t$. 

Note that the second term in Eq.~(\ref{am_12_18_23}) can be neglected for the following reason. $|\mathcal{E}_{\mu'}\rangle$ should be the same as $|\mathcal{E}_{\nu'}\rangle$ so that the integral can contribute to the density matrix of the KPO, however if they are the same $V_{\mu' \nu'}^{(\delta m')}=0$ and thus the second term becomes zero. The third term can also be neglected for the same reason.

\textcolor{black}{\subsection*{Probability $p_q$}}
Here, we follow the same method as Ref.~\cite{Silveri2017} to calculate $p_q$, which defines the probability of the normal-metal island state being $|q\rangle$. 
Because the elastic-tunneling rate is much larger than inelastic ones~\cite{Silveri2017},
we assume that $p_q$ can be determined by the elastic tunneling independently of the KPO state. 

The population is written as 
\begin{eqnarray}
p_q =  \frac{1}{Z} \prod_{q'=0}^{q-1} \frac{\Gamma_{q',m,m}^+}{\Gamma_{q'+1,m,m}^-},
\end{eqnarray}
where $Z$ is the normalization factor, and $\Gamma^\pm_{q,m,m}(V)$ is defined by
\begin{eqnarray}
\Gamma^\pm_{q,m,m}(V) = M_{m,m}^2\frac{R_K}{R_T}\sum_{\tau=\pm1} \overrightarrow{P}(\tau eV-E_q^\pm),
\label{Gamma_pm_1_30_24}
\end{eqnarray}
with $E_q^\pm = \frac{e^2}{2C_N}(1\pm 2q)$, $R_K=h/e^2$, and
\begin{eqnarray}
\overrightarrow{P}(E)=\frac{1}{h}\int_{-\infty}^\infty d\varepsilon n_s(\varepsilon) [1-f(\varepsilon)] f(\varepsilon-E).
\end{eqnarray}
In Eq.~(\ref{Gamma_pm_1_30_24}), $M_{m,m}^2$ is defined by
\begin{eqnarray}
M_{m,m}^2 = e^{-\rho_c}[L^0_m(\rho_c)]^2,
\end{eqnarray}
where $L^l_m(\rho_c)$ is the generalized Laguerre polynomials.
We also have $p_0 = 1/Z$ and $p_{-q} = p_q$. 
Equivalently $p_q$ is written as
\begin{eqnarray}
p_q =  \frac{1}{Z} \prod_{q'=0}^{q-1} \frac{\sum_{\tau=\pm 1}\overrightarrow{P}(\tau eV-E_{q'}^+)}{\sum_{\tau=\pm 1}\overrightarrow{P}(\tau eV-E_{q'+1}^-)}.
\end{eqnarray}

\begin{figure}
\begin{center}
\includegraphics[width=9.5cm]{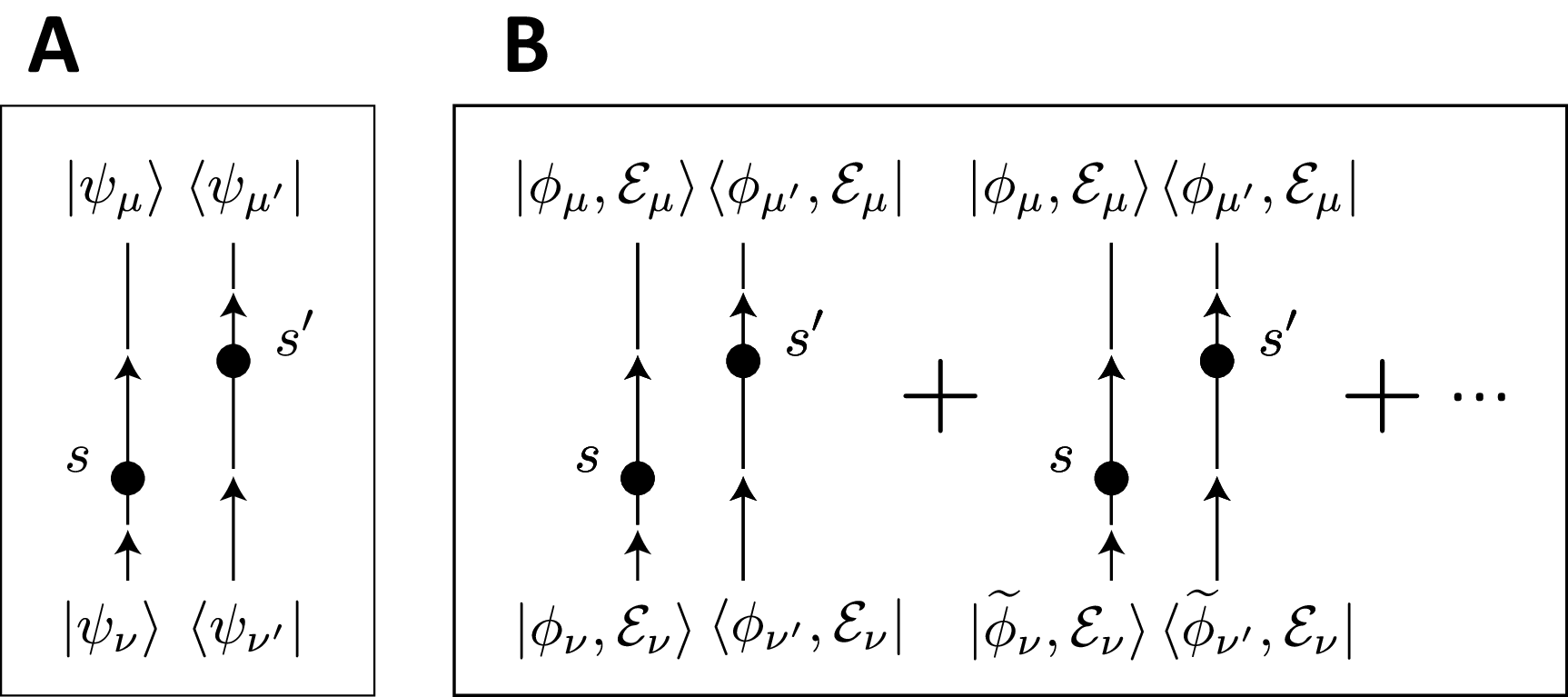}
\end{center}
\caption{
\textcolor{black}{({\bf A}) Schematic of the contribution from the integrand in Eq.~(\ref{integrand_12_10_24}) to $a_\mu(t)a_{\mu'}(t)$.
The left line represents $V_{\mu\nu}^{(\delta m)} e^{i(\omega_{\mu\nu} + \omega_{\rm RF} \delta m)s}a_\nu(0)$, while the right represents $ (V_{\mu'\nu'}^{(\delta m')})^\ast e^{-i(\omega_{\mu'\nu'} + \omega_{\rm RF} \delta m')s'}a_{\nu'}^\ast(0)$.
({\bf B}) The left pair of the lines is the same thing as panel (A), where we used $|\psi_\mu\rangle=|\phi_{\mu}, \mathcal{E}_\mu\rangle$, and $\mathcal{E}_\mu=\mathcal{E}_{\mu'}$ and $\mathcal{E}_\nu=\mathcal{E}_{\nu'}$.
The right pair is associated with the same environment states as the left, but different KPO states at $t=0$, $|\widetilde{\phi}_\nu\rangle$ and $|\widetilde{\phi}_{\nu'}\rangle$.  The KPO states corresponding to both the left and right pairs satisfy the first equation in Eq.~(\ref{matching_12_21_23}).
}
}
\label{interference_12_9_24}
\end{figure}

\textcolor{black}{\subsection*{Quantum interference effect associated with the level degeneracy}}
\textcolor{black}{
We explain the concept of the interference effect associated with the level degeneracy, with particular focus on the integrand of the time integrals in the third line of Eq.~(\ref{am_12_18_23}), 
\begin{eqnarray} 
V_{\mu\nu}^{(\delta m)} (V_{\mu'\nu'}^{(\delta m')})^\ast e^{i(\omega_{\mu\nu} + \omega_{\rm RF} \delta m)s} e^{-i(\omega_{\mu'\nu'} + \omega_{\rm RF} \delta m')s'}a_\nu(0)a_{\nu'}^\ast(0),
\label{integrand_12_10_24}
\end{eqnarray} 
where we replaced the second integral variable $s$ with $s'$ to distinguish it from the first. 
The contribution of this integrand to $a_\mu(t)a_{\mu'}^\ast(t)$ is schematically illustrated as Fig.~\ref{interference_12_9_24}(A).
The integrand can also be represented by the left pair of lines in Fig.~\ref{interference_12_9_24}(B),
as we consider the case where $|\mathcal{E}_\mu\rangle=|\mathcal{E}_{\mu'}\rangle$ and $|\mathcal{E}_\nu\rangle=|\mathcal{E}_{\nu'}\rangle$ in the calculation of the reduced density matrix for the KPO, as explained in the paragraph containing Eq.~(\ref{rho_KPO_12_11_24}).
This integrand can also be interpreted as the contribution from $\rho^{\rm KPO}_{\phi_\nu,\phi_{\nu'}}(0)$ to $\rho^{\rm KPO}_{\phi_\mu,\phi_{\mu'}}(t)$, because $a_\nu(0)a_{\nu'}^\ast(0)$ and $a_\mu(t)a_{\mu'}^\ast(t)$ are related to $\rho^{\rm KPO}_{\phi_\nu,\phi_{\nu'}}(0)$ and $\rho^{\rm KPO}_{\phi_\mu,\phi_{\mu'}}(t)$, respectively. 
The property of the time integral imposes a condition on the KPO states $\{|\phi_\nu\rangle,|\phi_{\nu'}\rangle\}$, as represented by the first line of Eq.~(\ref{matching_12_21_23}).
When there is the level degeneracy, multiple sets of such initial states exist, each with different KPO states $\{|\phi_\nu\rangle,|\phi_{\nu'}\rangle\}$ that satisfy Eq.~(\ref{matching_12_21_23}), but with the same environment states.
}
\textcolor{black}{
We refer to the contributions from such initial states, with different KPO states $\{|\phi_\nu\rangle,|\phi_{\nu'}\rangle\}$ but the same environment states, as the quantum interference effect in this paper.
Note that we term these contribution ``quantum interference" when the initial environment states are identical as in Fig.~\ref{interference_12_9_24}(B).
The contributions should not be called ``quantum interference" when the initial environment states are different, since there is no coherence between the different environment states.
}

\textcolor{black}{
This integrand is associated with $\Gamma^{(1)}(\phi_{\mu},\phi_{\mu'},\phi_\nu,\phi_{\nu'},V)$ in Eq.~(\ref{Gamma_12_21_23}).
The master equation in Eq.~(\ref{EOM_10_24_23}) accounts for the quantum interference via the summation with respect to the state of the KPO, $\phi_{\nu'}$, which satisfies Eq.~(\ref{matching_12_21_23}).
As a result, not only $\rho^{\rm KPO}_{\phi_0,\phi_1}(0)$ but also $\rho^{\rm KPO}_{\phi_1,\phi_0}(0)$ affects $\rho^{\rm KPO}_{\phi_0,\phi_1}(t)$ in our system where $\phi_0$ and $\phi_1$ are degenerate.
Especially, $\Gamma^{(1)}(\phi_{0},\phi_{1},\phi_1,\phi_{0},V)$ is significant to describe the bit-flip accurately.
If the contribution of $\rho^{\rm KPO}_{\phi_1,\phi_0}(0)$ is omitted by putting $\Gamma^{(1)}(\phi_{0},\phi_{1},\phi_1,\phi_{0},V)$ zero, the QCR-induced bit-flip rate becomes much larger than the correct one as shown in Fig.~\ref{coherence_2_26_24}(C).
}

\textcolor{black}{
Although, this paper considers the case that the two lowest energy levels are exactly degenerate, the formula of transition rates in Eq.~(\ref{Gamma_12_21_23}) remains approximately valid when the degeneracy is only approximate.
To discuss the validity of the formula in the presence of a small energy discrepancy, we consider the time integral $\int_0^t  ds e^{i(\omega_{\mu\nu} + \omega_{\rm RF} \delta m)s} 
 \int_0^t  ds e^{-i(\omega_{\mu'\nu'} + \omega_{\rm RF} \delta m')s}$ in the third line of Eq.~(\ref{am_12_18_23}), particularly focusing on the term including $c_{k\sigma}^\dagger d_{l\sigma}|q+1,\delta m+m\rangle \langle q,m|$ of $V^{(\delta m)}$ in Eq.~(\ref{V_12_22_23}) as an example.
The time integral can be considered as a function of $\varepsilon_l$, the energy of a quasiparticle in mode $l$, and this function exhibits two peaks, each with a width of $2\pi/\Delta t$ as explained in 
\textcolor{black}{subsection ``Time integrals in Eq.~(\ref{am_12_18_23})" in the Methods section}.
When the first equation in Eq.~(\ref{matching_12_21_23}) is satisfied, the two peaks overlap and the function works as a delta function for sufficiently large $\Delta t$.
If the energy difference between relevant levels is small enough compared to the width of the peaks, the levels can be considered  approximately degenerate, and Eq.~(\ref{matching_12_21_23}) is approximately satisfied.  
Assuming that $\Delta t$ is on the order of $0.1$~$\mu$s, the width of the peak is on the order of $2\pi \times 10$~MHz. Therefore, we consider that the formula in Eq.~(\ref{Gamma_12_21_23}) is approximately valid when the energy difference between relevant levels is on the order of 1~MHz or smaller. 
Here, we assumed that $\Delta t$ is on the order of $0.1$~$\mu$s so that the following conditions are satisfied.
The width of the peaks $2\pi/\Delta t$ is much smaller than the characteristic energy scale of the Fermi-Dirac distribution function, $k_BT_{N,S}/h \sim 2$~GHz, so that the function of $\varepsilon_l$ can be regarded as a delta function;
$\Delta t$ should be short enough so that the effect of the other decoherence sources can be neglected in the estimation of the QCR-induced transition rates. The second condition is represented as $\Delta t \ll 1/2\kappa \alpha^2$, where $2\kappa \alpha^2$ is the effective phase-flip rate caused by the single photon loss. 
Because the gap between the lowest levels and the first excited state is approximately 40~MHz in this study, the first excited level is considered apart enough from the lowest levels. 
} 



\section*{ACKNOWLEDGEMENTS}
We thank T. Ishikawa, M. A. Gal\'{i} Labarias, S. Kawabata, T. Yamamoto, H. Goto, and T. Kanao for useful discussions.
This paper is based on results obtained from a project, JPNP16007, commissioned by the New Energy and Industrial Technology Development Organization (NEDO), Japan.
TY acknowledges support from the JST Moonshot R\&D-MILLENNIA program (Grant No. JPMJMS2061).

\section*{AUTHOR CONTRIBUTIONS}
Conceptualization: SM. Methodology: SM. Investigation: SM. Validation: SK, TA, TY, AT. Supervision: SM. Writing-original draft: SM. Writing-review \& editing SM, SK, TA, TY, AT. 


\section*{CORRESPONDING AUTHOR}
Correspondence to Shumpei Masuda.

\end{document}